\documentclass[a4,12pt]{article}
\usepackage{graphicx}
\usepackage{subfigure}
\usepackage{amsmath}
\usepackage{amsfonts}
\usepackage{amssymb}
\usepackage{mathrsfs}
\usepackage{floatflt}
\usepackage[latin1]{inputenc}
\usepackage[T1]{fontenc}
\newtheorem{lemma}{Lemma}
\newtheorem{teorema}{Theorem}
\newtheorem{definizione}{Definition}

\newtheorem{corollario}{Corollary}

\def\min{{\rm min}}
 \def\eps{ { \varepsilon } }

\def\phi{{\varphi}}
\newcommand{\equal}{\buildrel {\rm def} \over {=} }
\newcommand{\vett}[1]{\mathbf{#1}}

\newcommand{\indice}{{\mathfrak{i}}}

\newcommand{\dif}{\mathrm{d}}

\newcommand{\Z}{{\mathbb Z}}

\newcommand{\M}{{\mathcal M}}
\newcommand{\I}{{\mathcal I}}

\newcommand{\resto}{{\mathcal R}}

\newcommand{\card}{\mathfrak n}
\newcommand{\N}{{\mathcal N}}

\newcommand{\Tc}{{\mathcal T}}

\newcommand{\Pc}{{\mathcal P}}

\newcommand{\C}{\mathbb{C}}

\DeclareMathOperator{\supp}{supp}
\DeclareMathOperator{\re}{Re}

\def\uno{{\kern+.3em {\rm 1} \kern -.22em {\rm l}}}

\title{Freezing of the optical--branch energy in a diatomic FPU chain
}
\author{Alberto Mario Maiocchi\thanks{Dipartimento di Matematica, Universit\`a degli Studi di Milano -- Via Saldini 50, I-20133, Milan, Italy. E-mail: \texttt{alberto.maiocchi@unimi.it} }}

\date{}
\begin{document}

\maketitle

\begin{abstract}
We study the exchange of energy between the modes of the optical
branch and
those of the acoustic one in a diatomic chain of particles, with
masses $m_1$ and $m_2$. We prove that, at small temperature and
provided  $m_1\gg m_2$, for the majority of the
initial data the energy of each branch is approximately constant for
times of order $\beta^{S/2}$, where $S=\lfloor \sqrt{m_1/m_2}/2\rfloor$ and
$\beta$ is the inverse temperature. The result is uniform in the
thermodynamic limit.
\end{abstract}

\section{Introduction}


In the present paper we study a variant of the Fermi--Pasta--Ulam
(FPU) chain, namely the Born and von K\'arm\'an lattice \cite{bornvk}
which consists of a chain of particles with nearest neighbour
interaction, having alternating heavy and light masses $m_1$ and
$m_2$. In this model the spectrum of the normal modes splits into two
separated branches called the acoustic and the optical branch. The
dynamics was studied numerically in \cite{GGMV} and a quite clear behaviour 
was observed: the energy of the optical branch
seems to be essentially an integral of motion, possibly also in the
thermodynamic limit. This could have a relevant
consequence for the thermodynamic behaviour of this model, so that a 
theoretical confirmation of this phenomenon seems to be in order.

In the present paper we actually prove that, uniformly in the
thermodynamic limit and for the majority of initial data,
the energy of the optical branch remains substantially constant over
times which increase as $\beta^{S/2}$, where
$S=\lfloor\sqrt{m_1/m_2}/2\rfloor$, with $m_1\gg m_2$, and where
$\beta$ is the inverse  
temperature. A detailed comparison of our result with numerical
observations will be provided at the end of Section~\ref{sez:principale}.

Our method of proof is a development of the ideas
introduced in \cite{C} and 
\cite{CM,RH13,primopasso}. Such tools 
allow one to implement perturbation theory for the majority of 
initial data, in a regime of interest for
statistical mechanics. 
In particular, in the paper \cite{primopasso} the original FPU model
was studied, showing that essentially any packet of harmonic
modes does not change significantly its energy for times of order
$\beta$. The obstruction to go to longer
times was due to the existence of small denominators.  As already
remarked in \cite{GGMV,BamGio}, the problem of small denominators does
not show up in studying the freezing of the optical
energy in the diatomic chain, so that the techniques developed in
\cite{primopasso} can be adapted to it. Here, however, the main difficulty is that
the form of the normal  
modes is more complicated and the estimates of the variances of the relevant
functions (see Lemma~\ref{lemma:stima} below) have to be rewritten from
scratch. Indeed, one has to work here on the Bravais lattice,
whose points are pairs of particles, and, moreover, one has to
introduce combinatorialtechniques based on the construction of
suitable graphs and trees. 

The problem we study is closely related that of equipartion of energy
in the original FPU model, in which all masses are equal, see
\cite{FPU}. Relevant contributions  were given by 
several authors, among which \cite{IC,rink,BGG04,BP,BCP13}; for a
recent review see \cite{GalFPU}. Further dynamical properties,
concerning the existence of localized solutions were
investigated both for the original FPU model (see
\cite{FPI,FPII,FPIII,FPIV,flach,Miz09})  and
the diatomic model considered here (see \cite{livi}). 

For the present model, the first analytic results on the freezing of
the optical energy were obtained in \cite{GGMV}, applying the
main theorem of \cite{vincoli2} to this model. In \cite{GGMV} the
authors provided a Nekhoroshev
type result, valid however for total energy $E$ smaller than some
inverse power of the number $N$ of degreees of freedom. In the
subsequent paper
\cite{BamGio} the authors introduced a suitable functional framework
which enabled them to prove energy 
freezing for any $N$, but still for a finite total
energy $E$, i.e., in a regime not relevant for statistical
mechanics. Instead, the result of the 
present paper  is uniform in the thermodynamic limit, in which $E/N$
remains constant as $N$ goes to infinity.

The description of the model and a precise statement of the result
is given in Section~\ref{sez:principale}, where the result on the
conservation of the energy of the optical (and also of the acoustic)
branch is given in 
Corollary~\ref{cor:principale}, which is deduced in a few lines from
the corresponding Theorem~\ref{teor:principale}, expressed in terms of
time correlations. In turn, Theorem~\ref{teor:principale} is a simple
consequence of the related Theorem~\ref{teor:ausiliario}, which
concerns the conservation of an auxiliary quantity. The latter Theorem, which
contains the main technical part of the work, is proved in
Section~\ref{sez:dimostrazione}, whereas in
Section~\ref{sez:corollario}  the deduction of
Theorem~\ref{teor:principale} and Corollary~\ref{cor:principale} is given.
A short discussion of the physical
consequences of the result is provided in the concluding
Section~\ref{sez:conseguenze}. 
A detailed
analysis of the normal modes of the system is given in
Appendix~\ref{app:modi}, while the remaining appendices contain some
technical Lemmas which have been isolated in order to clarify the
exposition of the proofs.

\section{Description of the model and main results}\label{sez:principale}

We consider a one--dimensional diatomic chain, constituted by two
species of masses $m_1$ and $m_2$ (with $m_1> m_2$). The Hamiltonian
of the system is 
$$
H=\sum_{j=1}^N\left(\frac{p_{j,1}^2}{2 m_1}+\frac{p_{j,2}^2}{2m_2}
\right) + \sum_{j=1}^N\left(V(x_{j,2}-x_{j,1})+
V(x_{j+1,1}-x_{j,2})\right) \ ,\quad N\gg 1\ ,
$$
in the canonically conjugated coordinates $p=\{p_{j,i}\}$,
$x=\{x_{j,i}\}$ in the phase space $\mathcal M=\mathbb R^{4N}$, where
$j=1,\ldots,N,$, $i=1,2$. The
potential $V$ corresponds to a nearest neighbour interaction, which we
assume of the form
$$
V(r)=\frac{K}2 r^2\left(1+ A r+Br^2\right) .
$$

As in the original work of  Born and von K\'arm\'an \cite{bornvk}, we impose
periodic boundary conditions, i.e., $\vett x_{N+j}=\vett x_j$ and
$\vett p_{N+j}=\vett p_j$, where we denote $\vett
x_j\equal(x_{j,1},x_{j,2})$ and $\vett
p_j\equal (p_{j,1},p_{j,2})$. Introducing the normal modes of the
system (see Appendix~\ref{app:modi}), the quadratic part of the
Hamiltonian takes the form 
$$
H_0=\frac12\sum_k\sum_{l=\pm}\left(\left|\hat p_k^l\right|^2+ \left. \omega^l_k
\right.^2 \left|\hat q_k^l\right|^2 \right) \ ,
$$
where $\hat p_k^\pm$, $\hat
q_k^\pm$, for $k=\lfloor- N/2 \rfloor +1,\ldots, \lfloor
N/2\rfloor$, are complex canonically conjugated variables.
This is a Hamiltonian of $2N$ harmonic oscillators with frequencies
\begin{equation}\label{eq:frequenze}
  \omega^\pm_k=\left(
  K\frac{m_1+m_2\pm\sqrt{m_1^2+m_2^2+2m_1m_2\cos(2\pi k/N)}}{m_1
    m_2}\right)^{1/2}\ .
\end{equation}
The index  $l=\pm$  splits the frequencies into two branches, which,
using a common terminology of solid state physics, will be called
optical branch 
(for $l=+$) and acoustic branch (for $l=-$). Notice that the
frequencies of the acoustic branch range from 0 to
$\sqrt{2K/m_1}$, while those of the optical branch
range from a minimum value $\sqrt{2K/m_2}$ to  
$\sqrt{2K/m_2}\cdot\sqrt{1+m_1/m_2}$. Thus, a
gap exists between the 
maximum value of the branch $\omega^-_k$ and the minimum value of the branch
$\omega^+_k$, and the gap increases with the ratio $m_1/m_2$. 
Consider the total energy  $E^+$ of the normal modes in the 
optical branch, and that of the acoustic branch, $E^-$,
namely
$$
E^+\equal\frac12\sum_k\left(\left|\hat p_k^+\right|^2+
\left. \omega^+_k 
\right.^2 \left|\hat q_k^+\right|^2 \right)\ ,\qquad E^-\equal
\frac12\sum_k\left(\left|\hat p_k^-\right|^2+ 
\left. \omega^-_k 
\right.^2 \left|\hat q_k^-\right|^2 \right)\ .
$$

In order to formulate precisely this statement (see
Theorem~\ref{teor:principale}), we introduce the Gibbs measure in phase space $\M=\mathbb
R^{2N}\times \mathbb R^{2N}$, namely, 
$$
 \mu(\dif p\,\dif x)\equal\frac{\exp(-\beta H(p,x))}{Z(\beta)} \dif p\, \dif
x\ ,
$$
where $\dif p\,\dif x$ denotes the Lebesgue measure $\prod_j\dif
p_{j,1}\dif p_{j,2}\dif x_{j,1}\dif x_{j,2}$, while  $\beta>0$ is the
inverse temperature and $Z(\beta)\equal\int_\M \exp(-\beta H(p,x))
\dif p\, \dif x$ the partition function. It is well known that $\mu$
is invariant for the flow. For any 
dynamical variable $F$, the mean  $\langle F\rangle$ and the variance
$\sigma^2_F$ are thus defined by
$$
\langle F\rangle \equal \int_\M F \dif \mu\ , \qquad
\sigma^2_F\equal \langle \left(F - \langle F\rangle\right)^2\rangle\ .
$$
The time autocorrelation $C_F(t)$ of $F$ is defined by
$$
C_F(t)\equal \langle F_t\, F\rangle- \langle F\rangle^2 \ ,
$$
in which $F_t=F\circ g^t$ and $g^t$ is the flow generated by $H$.


In the spirit of the statistical approach pursued here, the result on
the conservation of the energies $E^+$ and $E^-$ of the optical and
the acoustic branch is naturally stated in terms of their
correlations
\begin{teorema}\label{teor:principale}
There exist constants $\beta^*>0$, $ 
N^*>0$, $M>2$ and $K_1,K_2>0$ such that, for any $\beta>\beta^*$, $N>
N^*$ and for any value of  $m_1/m_2>M$, the following bounds hold
\begin{equation}\label{eq:alti}
\left|C_{E^\pm}(t)-C_{E^\pm}(0)\right| \le K_2\left(\frac {1}{\sqrt
  \beta} + \frac{m_2}{m_1}\right)\sigma^2_{E^\pm} \ , \quad \mbox{for } t\le
K_1\beta^{S/2} \ ,
\end{equation}
where $S=\lfloor \sqrt{m_1/m_2}/2\rfloor$.
\end{teorema}
\begin{corollario}\label{cor:principale}
There exist a measurable set $\mathcal J$ and $C>0$ such that  $\mu({\mathcal
  J}^c)<C\beta^{-S/2}$ and
\begin{equation*}
  \frac{\left| E^\pm(g^t x) -E^\pm(x)\right|}{\sigma_{E^\pm}}\le
  C\left(\frac{1}{\sqrt\beta}+\frac{m_2}{m_1}\right) \ , \quad
  \mbox{for } |t|\le C\beta^{S/2}\ ,
\end{equation*}
if $x\in\mathcal J$.
\end{corollario}

It is interesting to compare our result with the numerical
observations of \cite{GGMV}. In that paper the authors measured some
average of $\tfrac{d}{dt} E^+$ bounding this quantity by
$$
A(N)\frac{m_2}{m_1}\exp\left(-B\frac{m_1}{m_2}\right)\ ,
$$ with a constant $A(N)$ which diverges less than logarithmically as
$N$ increases and $B$ independent of $N$. We remark that the small
parameter of \cite{GGMV} is $m_2/m_1$. Furthermore the initial data
considered in \cite{GGMV} did not have small specific energy: only the
specific energy present in the optical branch was assumed to be small.
Our result is somehow stronger than the one observed in \cite{GGMV}
since it is completely uniform with $N$. On the other hand our small
parameter is the temperature, so that we are studying a more
particular regime.


\section{Proof of Theorem~\ref{teor:principale}: main technical part}
\label{sez:dimostrazione}
The proof is performed by formally constructing a constant of motion
$\Phi$ through a formal series expansion starting from
$$
\Phi_0=\sum_k \frac{1}{2\omega^+_k} \left(\left|\hat p_k^+
\right|^2+\left. \omega^+_k \right.^2 \left|\hat q_k^+\right|^2
\right)\ ,
$$
which is the sum of the actions of the modes in the optical branch.
The series is then truncated at a given order $S$, and it is
shown that $S$ can be so chosen that the time autocorrelation of the
truncated quantity $\Phi^{(S)}$ has small variation over long
times. Indeed, the main technical part of the present work can
be summarized in the following
\begin{teorema}\label{teor:ausiliario}
There exist a polynomial $\Phi^{(S)}$ of degree $S=\lfloor
\sqrt{m_1/m_2}/2\rfloor$ and constants $\beta^*>0$, $ 
N^*>0$, $M>2$ and $K_1,K_2>0$ such that, for any $\beta>\beta^*$, $N> N^*$
and for any value of 
$m_1/m_2>M$, 
\begin{equation}\label{eq:ausiliario}
\sigma_{\dot{\Phi}^{(S)}} \le \frac{K_1}{\beta^{S/2}}
\sigma_{\Phi^{(S)}}\ , \qquad
\sigma_{\Phi^{(S)}-\Phi_0}\le\frac{K_2}{\sqrt\beta}\sigma_{\Phi_0} \ . 
\end{equation}
\end{teorema}
The proof of Theorem~\ref{teor:principale} easily follows,
through standard arguments, as shown in the following
Section~\ref{sez:corollario}.

The rest of the present section is devoted to the proof of
Theorem~\ref{teor:ausiliario}. We briefly 
illustrate first the formal construction scheme for 
the integral of motion in section~\ref{sez:formale}. In order to give
quantitative estimates, in section~\ref{sez:stime} we define the
classes of functions with which we have to deal, which are actually
suitable polynomials, and construct a sequence of Banach spaces
$\Pc_s$ of homogeneous polynomials of degree $s$, with a suitable
norm. This is basically an adaptation of the techniques of
\cite{giorgilli} to our class of polynomials, with the adoption of
some tools from \cite{GPP13}. The relation between the norms $\Pc_s$
and the variances with respect to the Gibbs measure, which are the ones we are
interested in, is displayed in the following
section~\ref{sez:varianze}. Here (and in the related
Appendix~\ref{app:scambiabili}) is contained the main technical novelty
of the work, namely, a complete reformulation and extension of the
techniques introduced in \cite{primopasso} to control the relation
between the norms, based on a careful counting of the terms entering
the variances, through the introduction of suitable graphs and trees. In section~\ref{sez:fine_dimostrazione}
the final estimates are summed up and the proof of
Theorem~\ref{teor:ausiliario} is completed.

\subsection{The formal construction scheme for the constant of motion}
\label{sez:formale}
We construct a formal integral  of motion $\Phi$ by using the algebraic
algorithm involving Lie transforms which was presented in
\cite{giorgilli}. First, given a generating sequence
$\chi=\{\chi_s\}_{s\ge 1}$, consider the formal linear operator
$T_\chi$, acting on formal polynomials, defined by
$$
T_\chi=\sum_{s\ge 0} E_s\ , \quad\mbox{where } E_0=\mathbb I\ , \quad
E_s=\sum_{j=1}^s\frac js L_{\chi_j}E_{s-j}\ , 
$$
in which $L_{\chi_j}\cdot = \{\chi_j,\cdot\}$ and $\{\cdot,\cdot\}$
denotes Poisson brackets.


The sequence $\chi_j$, in turn, is determined in the following way. Expand
the Hamiltonian in homogeneous polynomials,
$H=\sum_{s\ge 0}H_s$, with $H_s$
homogeneous polynomials of degree $s+2$ in the canonical
coordinates. The functions $\chi_s$ are then determined recursively by
solving an equation of the form
\begin{equation}\label{eq:omologica}
 L_0\chi_s = Z_s-\Psi_s\ ,
\end{equation}
where $L_0=L_{H_0}$, $\Psi_s$ is given and $Z_s$ is a normal form,
that must commute with $\Phi_0$, or, equivalently, with the resonant
part of the Hamiltonian
$$
H_\Omega=\Omega\sum_k\frac{\left|\hat p_k^+
\right|^2+\left. \omega^+_k \right.^2 \left|\hat q_k^+\right|^2
}{2\omega_k^+}=\Omega \Phi_0\ , 
$$
with
$\Omega$ maximum optical frequency, i.e.,
$$
\Omega=\omega_0^+=\sqrt{\frac{2K(m_1+m_2)}{m_1m_2}}\ .
$$

One of the main points is the construction of $Z_s$ and $\chi_s$
solving (\ref{eq:omologica}). Recall first that any
polynomial $\Psi_s$ can be decomposed into a kernel and a range component of the
operator $L_\Omega=L_{H_\Omega}$. Denote by $\Pi_{\mathcal N}$ and $\Pi_{\mathcal
  R}$ the corresponding projections. We define $Z_s=\Pi_{\mathcal
  N}\Psi_s$ and then solve through Neumann formula (see
\cite{BPC07,GPP13}) 
\begin{equation}\label{eq:omologica_inversa}
L_0\chi_s=\left(L_\Omega+L_{H_0-H_\Omega}\right)\chi_s=
\Pi_{\mathcal R} \Psi_s\ .
\end{equation}
The quantity $\Psi_s$ is recursively defined by the formula
\begin{equation*}
  \begin{split}
    \Psi_1&=H_1\ ,\\
    \Psi_s& =H_s +\sum_{l=1}^{s-1}\frac ls L_{\chi_l} H_{s-l}
    +\sum_{l=1}^{s-1}\frac ls E_{s-l}Z_l\  , s\ge 2\ ,
    \end{split}
\end{equation*}
and $\chi_s,Z_s$ are the solutions of the homological equation
(\ref{eq:omologica_inversa}). Then, by considering 
$$
\Phi=T_\chi \Phi_0=\sum_{j\ge 0}\Phi_j \quad \mbox{and }
\Phi^{(S)}\equal \sum_{j=0}^S \Phi_j\ ,
$$
the theory of \cite{giorgilli} ensures that
$$
\dot{\Phi}^{(S)}\equal
\left\{\Phi_S,H_1+H_2\right\}+\left\{\Phi_{S-1},H_2\right\}\ ,
$$
which is the formula to be used for the quantitative estimates.

\subsection{Definition of the class of polynomials and quantitative estimates}
\label{sez:stime}
We start with a further (standard) change of variables that makes the operator
$L_0$ diagonal:
\begin{equation}\label{eq:trasf_xi_q}
\xi_k^\pm=\frac{\hat p_k^\pm+ i\omega^\pm_k\hat
  q^\pm_{-k}}{\sqrt{2}}\ , \quad 
\eta_k^\pm=\frac{\hat p^\pm_{-k}- i\omega^\pm_k\hat
  q^\pm_k}{\sqrt{2} }\ .
\end{equation}
This transformation brings to the Poisson brackets
$\{\xi^l_k,\eta^{l'}_{k'}\}=i\omega_k^\pm\delta_{k,k'}\delta_{l,l'}$
and gives the quadratic Hamiltonian the form
$H_0=\sum_k\sum_\pm  \xi_k^\pm \eta_k^\pm$.

In order to define the class of polynomials we will meet, we start by
introducing the monomials 
$$
\Xi^s_{\sigma,k,l} \equal{\xi^{l_1}_{k_1}}^{(1+\sigma_1)/2 }
{\eta^{l_2}_{k_1}}^{(1 -\sigma_1)/2}\ldots 
{\xi^{l_s}_{k_s}}^{(1+\sigma_s)/2}{\eta^{l_s}_{k_s}}^{(1-\sigma_s)/2}
\ ,\quad s\geq 3\ ,
$$
which have degree $s$, where
\begin{equation}
  \begin{split}
\label{indici.1}
\sigma&=(\sigma_1,\ldots,\sigma_{s})\ ,\ \sigma_j=\pm1\ ,\\
k&=(k_1,\ldots,k_s)\ ,\ k_j=\lfloor-N/2\rfloor+1,\cdots,\lfloor N/2\rfloor\ ,\\
l&=(l_1,\ldots,l_s)\ ,\ l_j=\pm\ ,
\end{split}
  \end{equation}
and observe that a fundamental property of all monomials 
is that the indices $k$ have  a relation of the form
\begin{equation}
\label{sel}
\tilde \tau\cdot k= nN\ ,
\end{equation}
for some
\begin{equation}
\label{indici3}
\tilde \tau=(\tilde \tau_1,\ldots,\tilde \tau_s)\ ,\quad \tilde
\tau_l=\pm1\quad n=\lfloor-(s-1)/2\rfloor,\ldots,\lfloor(s-1)/2\rfloor .
\end{equation}
In the following we will denote by  $\I_s$ the set of indices
$(\sigma,\tilde \tau,k,l,n)$ of the form \eqref{indici.1},
(\ref{indici3}).

\begin{definizione}
\label{def.P}
We say that $f\in \Pc_s$ if it can be written as
\begin{equation}
\label{ps}
f=\frac 1{N^{(s-2)/2}}\sum_{(\sigma,\tilde \tau,k,l,n)\in \I_s}
f_{\sigma,\tilde \tau,l,n}\left(\frac{k_1}{N},\ldots,\frac{k_s}{N}\right)
\Xi^s_{\sigma,k,l} \delta^n_{\tilde \tau\cdot k}\ ,
\end{equation}
where $f_{\sigma,\tilde \tau,l,n}:[0,1]^s\to\C$ are continuous
functions and $\delta^n_j$ is a shortcut for the
Kronecker delta $\delta_{j,nN}$.
\end{definizione}


In $\Pc_s$ we define the norm
\begin{equation}
\label{norma.f}
\left\|f\right\|_+\equal \max_{(\sigma,\tilde \tau, k,l,n)\in \I_s}
\left|f_{\sigma,\tilde \tau,n}\left(\frac{k_1}{N},\ldots,\frac{k_s}{N}
\right)\right|\delta_{\tilde \tau\cdot k}^n\ .
\end{equation}

One has the lemma (proved in Appendix~\ref{app:par_poisson})
\begin{lemma}\label{lemma:par_poisson}
 If $f\in \Pc_s$, $g\in
  \Pc_r$, then $\{f,g\}\in\Pc_{r+s-2}$. Moreover, one has
$$
\left\|\{f,g\}\right\|_+ \le 2^4 \Omega\, r\,s\,\min(r,s)\, \left\|f\right\|_+
\left\|g\right\|_+\ .
$$
\end{lemma}

We now have all the tools needed in order to construct the solutions
for the homological equation \eqref{eq:omologica}: we intend to show,
in a way completely analogous to \cite{GPP13}, to which the reader
will be referred for some proofs, that in our case
eq.~\eqref{eq:omologica} can be solved for $s\le S=S(m_1/m_2)$, with
$\chi_s, Z_s$ and $\Psi_s$ belonging to the Banach spaces  $\Pc_{s+2}$. 

First, we point out that the monomials $\Xi^s_{\sigma,k,l}$ are
eigenfunctions for the operators  $L_0$ and $L_\Omega$, with
eigenvalues given by
\begin{equation}\label{eq:autovalori_operatori}
L_0\Xi^s_{\sigma,k,l}=i\left(\sum_{j=1}^s\sigma_j \omega_{k_j
}^{l_j}\right) \Xi^s_{\sigma,k,l} \ , \quad
L_\Omega\Xi^s_{\sigma,k,l}=i\Omega\left(\sum_{j=1}^s\sigma_j\delta_{l_j,+}
\right) \Xi^s_{\sigma,k,l} \ .
\end{equation}
For this reason, both $L_0$ and $L_\Omega$ map $\Pc_s$ in itself and,
in particular, $\Pc_s$ is the direct sum of $\N_s$ and
$\resto_s$, the kernel and the range of
$L_\Omega$, respectively. Since $L_0=L_\Omega+L_{\Theta_0}$, where
$$
\Theta_0=\sum_k\left(\left(1-\frac{\Omega}{\omega_k^+}\right)
\frac{\left|\hat p_k^+   
\right|^2+\left. \omega^+_k \right.^2 \left|\hat q_k^+\right|^2}2 +
\frac{\left|\hat p_k^- 
\right|^2+\left. \omega^-_k \right.^2 \left|\hat q_k^-\right|^2}2
\right) \ ,
$$
we note
then that
$$
L_0^{-1} = \left(\mathbb I +K\right)^{-1}L_\Omega^{-1}\ , \quad
\mbox{with } K\equal L_\Omega^{-1}L_{\Theta_0}
$$
and that $K:\resto_s\mapsto \resto_s$, because $L_{\Theta_0}f\in
\resto_s$, if $f\in\resto_s$, as it can be shown in virtue of the
Jacobi identity and of the fact that $\{\Theta_0,H_\Omega\}=0$
(cfr. Lemma~4.1 in \cite{GPP13}). The operator $L_0$ can be then
inverted on $\resto_s$, by using Neumann formula, which holds provided
 $\|K\|<1$ on $\resto_s$.

This ensures that a solution for the homological equation 
(\ref{eq:omologica}) up to a given order can be constructed, as is
expressed by the following lemma, whose proof is deferred to
Appendix~\ref{app:omologico}:
\begin{lemma}\label{lemma:omologico}
Let $S=\lfloor \sqrt{m_1/m_2}/2\rfloor$. Then for $s\le S$ we have that
$\Psi_s\in \Pc_{s+2}$ and $K:\resto_s\mapsto \resto_s$, with $\|K_{\resto_s}\|\le
1/2$ on $\resto_s$. Moreover, for $s\le S$ there exist $Z_s, \chi_s,\in
\Pc_{s+2}$ such that: 
\begin{enumerate}
\item they are solutions for \eqref{eq:omologica};
\item $Z_s$ is in involution with $H_\Omega$, i.e., $Z_s\in \N_{s+2}$;
\item If  $\|H_s\|_+\le B^ss!$ there exists $C>0$ such that,
  for $ 1\le s\le S$,
  \begin{equation}\label{eq:induzione1}
    \left\|Z_s\right\|_+\le   \left\|\Psi_s\right\|_+ \le
    B^sC^{s-1}s!\ ,
  \end{equation}
\item for  $f_l\in\Pc_{l+2}$ and $1\le s\le S$, one has
  $E_sf_l\in  \Pc_{s+l+2}$,  with
  \begin{equation}\label{eq:induzione2}
    \begin{split}
\left\|E_sf_l\right\|_+&\le \frac 14B^sC^s
\frac{(s+l)!}{l!}\left(\frac{1}{s!}+\frac{1}{l+1}\right)
\left\|f_l\right\|_+\ , \quad \mbox{for }l\ge 1\ ,\\ 
\left\|E_sf_l\right\|_+&\le \frac 14B^sC^s
{(s+1)!}\left\|f_l\right\|_+\ , \quad \mbox{for } l=0\ .
    \end{split}
  \end{equation}  
\end{enumerate}
\end{lemma}

\subsection{Estimate for the variances}\label{sez:varianze}
The main result of this section is that, for any $f\in \Pc_s$, its
variance can be bounded from above by the following
\begin{lemma}\label{lemma:stima} There exist $N_0>0$ and $C>0$ such
  that, for any $2\le s\le S$, for any $N>N_0$ and any 
$f\in\Pc_s$, one has
  \begin{equation*}
\sigma^2_f\le N \frac{C^{2s}}{\beta^{s}}(2s!)^{3/2} \left\| f\right\|^2_+\ .
\end{equation*}
\end{lemma}

\noindent
Proof. By the definition of  variance and that of the class $\Pc_s$
one has
\begin{equation}\label{eq:varianza_polinomi}
  \begin{split}
\sigma^2_f=& \frac{1}{N^{s-2}}\sum_{\substack{(\sigma,\tilde \tau,k,l,n)\in
  \I_s\\(\sigma',\tilde \tau',k',l',n')\in
  \I_s}} f_{\sigma,\tilde \tau,l,n}\,f_{\sigma',\tilde \tau',l',n'}
\delta^n_{\tilde \tau\cdot k} \delta^{n'}_{\tilde \tau'\cdot k'}  \\
& \quad\quad \times\left( \langle
\Xi^s_{\sigma,k,l}\Xi^s_{\sigma',k',l'}\rangle -  \langle 
\Xi^s_{\sigma,k,l}\rangle \langle\Xi^s_{\sigma',k',l'}\rangle \right) \\
\le & \frac{1}{N^{s-2}} \left\|f\right\|_+^2\\
&\left(\sum_{\substack{(\sigma,\tilde \tau,k,l,n)\in
  \I_s\\(\sigma',\tilde \tau',k',l',n')\in
  \I_s}}
\delta^n_{\tilde \tau\cdot k} \delta^{n'}_{\tilde \tau'\cdot k'}\left| \langle
\Xi^s_{\sigma,k,l}\Xi^s_{\sigma',k',l'}\rangle -  \langle 
\Xi^s_{\sigma,k,l}\rangle \langle\Xi^s_{\sigma',k',l'}\rangle
\right|\right) \ .
  \end{split}
\end{equation}
The main part of the proof is then to show that the sum in the last
line can be bounded from above by
$N^{s-1}C^{2s}(2s!)^{3/2}/\beta^s$. This seems quite difficult, and in
particular the dependence on $N$ seems to pose a big problem: note, in
fact, that the sum over $k$ and $k'$, taking into account the
constraint imposed by the Kronecker deltas, contains a number of
terms of order $N^{2s-2}$. A huge number of terms in the sum must thus
vanish, in order to reduce the size, precisely as many as would vanish 
if  $\Xi^s_{\sigma,k,l}$ and $\Xi^s_{\sigma',k',l'}$ were uncorrelated
for $k\neq k'$. This is not the case, but it can be proved that the
correlation between the two monomials is always zero, unless the
components $k$ and $k'$ satisfy some linear relations, which will be
expressed by the introduction of some suitable Kronecker deltas,
as we detail now.

Fix a positive integer $R$, and consider the vectors
$\tau=(\tau_1,\ldots,\tau_R)$, with the $j$--th component
$\tau_j=0,\pm 1$. Denote by $\Z_3^R$ the set of such vectors and by
$\supp(\tau)$ the  set of indices $j$ such that $\tau_j\neq 0$.
\begin{definizione}
A collection $\tau^{(1)},\ldots,\tau^{(S_1)}$ of vectors
$\tau^{(i)}\in \Z_3^R$ will be said $R$--admissible, or simply
admissible, if $S_1\le R$, the supports $\supp(\tau^{(i)})$
constitute a partition of the set $\{1,\ldots,R\}$ in disjoint subsets
and if 
$$
\min(\supp(\tau^{(i)}))< \min(\supp(\tau^{(j)})) \Longleftrightarrow
i<j\ .
$$
\end{definizione}
We will denote by $\Tc_{R}$ the set of $R$--admissible vectors;
the introduction of this class enables us to state the following
lemma, which comes from the fact that  $(\vett p_j,\vett  r_j)$, with
$\vett r_j\equal(x_{j,2}-x_{j,1},x_{j,1}-x_{j-1,2})$, are
exchangeable variables (see Appendix~\ref{app:scambiabili}, where the
proof of this Lemma is reported):
\begin{lemma}\label{lemma:somma_scambiabili}
For any $S_1<s+s'$ there exist constants
 $c^{(\tau^{(1)},\ldots,\tau^{(S_1)})}_s>0$, independent of $k,k'$ and
$N$, such that
 \begin{equation}\label{eq:correlazione_monomi}
    \begin{split}
\left| \langle
\Xi^s_{\sigma,k,l}\Xi^{s'}_{\sigma',k',l'}\rangle -\right. &\left. \langle 
\Xi^s_{\sigma,k,l}\rangle \langle\Xi^{s'}_{\sigma',k',l'}\rangle
\right| \le \sum_{S_1=1}^{s+s'}N^{S_1-(s+s')/2}\\
&\sum_{(\tau^{(1)},\ldots, \tau^{(S_1)}) \in
  \Tc_{s+s'}}\sum_{n_1,\ldots n_{S_1}}  \delta^{n_1}_{\tau^{(1)}\cdot K}
\cdots \delta^{n_{S_1}}_{\tau^{(S_1)}\cdot K}
c^{(\tau^{(1)},\ldots,\tau^{(S_1)})}_{s,s'}\ ,
    \end{split}
    \end{equation}
in which $K=(k_1,\ldots,k_s,k'_1,\ldots,k'_{s'})$.
\end{lemma}

\noindent
    {\bf Remark:} In each sum over $n_i$ the terms
    $\delta^{n_i}_{\tau^{(i)}\cdot K}$ can be different from zero only
    for $n_i$ ranging from $\lfloor -\mathfrak n_i/2\rfloor +1$ to
    $\lfloor \mathfrak n_i/2\rfloor $, where $\mathfrak n_i$ is the
    cardinality of $\supp \tau^{(i)}$. This because $\tau^{(i)}\cdot
    K$ ranges from $\mathfrak n_i(\lfloor -N/2\rfloor+1)$ to
    $\mathfrak n_i\lfloor N/2\rfloor$.

We come back to the estimate of the variance and insert
(\ref{eq:correlazione_monomi}) into (\ref{eq:varianza_polinomi}),
observing that
\begin{equation}
  \begin{split}
\sigma^2_f\le&  
\left\|f\right\|_+^2\sum_{n,n'}\sum_{k,k'}\delta^n_{\tilde \tau\cdot
  k} 
\delta^{n'}_{\tilde \tau'\cdot k'} \sum_{S_1=1}^{2s}N^{S_1+2-2s}\\
&\sum_{(\tau^{(1)},\ldots, \tau^{(S_1)}) \in
  \Tc_{2s}}\sum_{n_1,\ldots n_{S_1}}  \delta^{n_1}_{\tau^{(1)}\cdot K}
\cdots \delta^{n_{S_1}}_{\tau^{(S_1)}\cdot K}
c^{(\tau^{(1)},\ldots,\tau^{(S_1)})}_{s,s}\ .
  \end{split}
  \end{equation}
The Kronecker deltas $\delta^{n_i}_{\tau^{(i)}\cdot K}$ represent some
linear relations that the vector $K=(k,k')$ has to satisfy, relations
which are all independent, because the supports of $\tau^{(i)}$ are
disjoint. Thus, in the sum over $(k,k')$ only
$2s-S_1$ independent terms are left. In the general case, it cannot be
proved that the further constraints imposed by $\delta^{n}_{\tilde
  \tau\cdot k}$ and $\delta^{n'}_{\tilde \tau'\cdot  k'}$ entail
another independent linear restriction on the sum, but this certainly
happens outside a set of indices
$(\tau^{(1)},\ldots,\tau^{(S_1)})$ which we now specify. Consider the
collection $\bar \Tc\subset \Tc_{2s}$ of $(\tau^{(1)},\ldots,\tau^{(S_1)})$ 
such that, for any  $\tau^{(i)}$, either
$\supp \tau^{(i)}\subset \{1,\ldots,s\}$, or  $\supp 
\tau^{(i)}\subset \{s+1,\ldots,2s\}$, and denote by $\bar \Tc^c$ its
complement in $\Tc_{2s}$. Then it can be shown that (as is
proved in Lemma~9 of \cite{primopasso}), for 
$(\tau^{(1)},\ldots,\tau^{(S_1)})\in\bar \Tc^c$, at least one among 
$\delta^{n}_{\tilde \tau\cdot k}$ and $\delta^{n'}_{\tilde \tau'\cdot
  k'}$ implies a constraint on the sum over $(k,k')$ which is
independent of those imposed by $\delta^{n_i}_{\tau^{(i)}\cdot K}$.

Coming to formulas, this means that, since
$$
\sum_{\buildrel{k_i}\over{i\in
    \supp(\tau^{(i)})}}\sum_{n_i}\delta^{n_i}_{\tau^{(i)}\cdot K} \le
\mathfrak n_i N^{\mathfrak n_i-1}\ ,
$$
if $N>s$, one has
$$
\sum_{k,k'}\sum_{n,n'}\delta^n_{\tilde \tau\cdot
  k}  \delta^{n'}_{\tilde \tau'\cdot k'}\sum_{n_1,\ldots n_{S_1}}  \delta^{n_1}_{\tau^{(1)}\cdot K}
\cdots \delta^{n_{S_1}}_{\tau^{(S_1)}\cdot K} \le \prod_{i=1}^{S_1}
\mathfrak n_iN^{\mathfrak n_i-1}=N^{2s-S_1}\prod_{i=1}^{S_1}
\mathfrak n_i\ .
$$
If, moreover,  $(\tau^{(1)},\ldots,\tau^{(S_1)})\in\bar \Tc^c$, on
account of Lemma~9 of \cite{primopasso} the estimate can be refined with
$$
\sum_{k,k'}\sum_{n,n'}\delta^n_{\tilde \tau\cdot
  k}  \delta^{n'}_{\tilde \tau'\cdot k'}\sum_{n_1,\ldots n_{S_1}}  \delta^{n_1}_{\tau^{(1)}\cdot K}
\cdots \delta^{n_{S_1}}_{\tau^{(S_1)}\cdot K} \le N^{2s-S_1-1}s\prod_{i=1}^{S_1}
\mathfrak n_i\ .
$$
We can then write
\begin{equation*}
  \begin{split}
\sigma^2_f\le N  \left\|f\right\|_+^2\sum_{S_1=1}^{2s}
&\Biggl(s\sum_{(\tau^{(1)},\ldots, \tau^{(S_1)}) \in
  \bar \Tc^c}\mathfrak n_1\cdots\mathfrak n_{S_1}
c^{(\tau^{(1)},\ldots,\tau^{(S_1)})}_{s,s}\\
&+N\sum_{(\tau^{(1)},\ldots,
  \tau^{(S_1)}) \in 
  \bar \Tc}\mathfrak n_1\cdots\mathfrak n_{S_1}
c^{(\tau^{(1)},\ldots,\tau^{(S_1)})}_{s,s}\Biggr) \ .
  \end{split}
\end{equation*}

This is enough for our aims, since in our case
(see Appendix~\ref{app:scambiabili}) a precise estimate of the
constants $c$ entering the previous formula is available:
\begin{lemma}\label{lemma:indipendenti}
There exists $C>0$ such that
\begin{equation*}
  \begin{split}
\sum_{S_1=1}^{2s}\sum_{(\tau^{(1)},\ldots, \tau^{(S_1)})\in  \bar \Tc^c}
\mathfrak n_1\cdots\mathfrak n_{S_1}
c^{(\tau^{(1)},\ldots,\tau^{(S_1)})}_{s,s}\le  C^{2s}\frac {(2s!)^{3/2}}{\beta^s}\ .
\\
\sum_{S_1=1}^{2s}\sum_{(\tau^{(1)},\ldots, \tau^{(S_1)})\in  \bar \Tc}
\mathfrak n_1\cdots\mathfrak n_{S_1}
c^{(\tau^{(1)},\ldots,\tau^{(S_1)})}_{s,s}\le \frac 1N C^{2s}\frac {(2s!)^{3/2}}{\beta^s}\ ,
  \end{split}
\end{equation*}
\end{lemma}

The thesis of lemma~\ref{lemma:stima} follows then easily by applying
this estimate.

\subsection{Conclusion of the proof of Theorem~\ref{teor:ausiliario}}
\label{sez:fine_dimostrazione}
In virtue of Lemma~\ref{lemma:omologico}, we can construct approximants
of the first integral $\Phi$ as $\Phi^{(r)}=\sum_{s= 0}^r \Phi_s$,
with $\Phi_s=E_s\Phi_0\in \Pc_{s+2}$ and $\left\|\Phi_s\right\|_+\le
s!C^s$. This can be done for any $r\le S=\lfloor \sqrt{m_1/m_2}/2\rfloor$.

Since the variables $\vett p_j$ are independent of the
variables $\vett x_j$, it is easy to show that
\begin{equation}\label{eq:minorazione_Phi_0}
\sigma_{\Phi_0}\ge \sigma_{\sum_k|p_k^+|^2/2\omega_k^+}\ge
\frac{\sqrt N C_1}{\beta}\ ,  
\end{equation}
while
\begin{equation}\label{eq:maggiorazione_differenza}
\sigma_{\Phi^{(S)}-\Phi_0}\le \sum_{s=1}^S \sigma_{\Phi_s}\le \sqrt N\sum_{s=1}^S
\frac{C_2^s}{ \beta^{(s+2)/2}} (s!)^{5/2}\ ,
\end{equation}
because of Lemma~\ref{lemma:stima}. Hence follows
\begin{equation}\label{eq:minorazione_Phi_s}
\sigma_{\Phi^{(S)}}\ge \sigma_{\Phi_0}- \sigma_{\Phi^{(S)}-\Phi_0} \ge
\frac{\sqrt N C_3}{\beta}\left(1-\sum_{s=1}^S\frac{C_3^s}{
  \beta^{s/2}} (s!)^{5/2}\right)\ ,
\end{equation}
and, for $\beta$ large enough, that
$$
\sigma_{\Phi^{(S)}-\Phi_0}\le\frac{K_2}{\sqrt\beta} \sigma_{\Phi_0}\ ,
$$
i.e., the second statement of (\ref{eq:ausiliario}).

In order to estimate the derivative with respect to the flow of
$\Phi^{(S)}$, as already remarked we point out that (see \cite{giorgilli}) this is equal to
$$
\dot \Phi^{(S)}= \sum_{s=0}^S\{\Phi_s,\sum_{s'\ge S-s+1}H_{s'}\}\ .
$$
In our case, in which $H_s=0$ for $s\ge 3$, we have to estimate
$$
\Upsilon_S=\{\Phi_S,H_1\}+\{\Phi_{S-1},H_2\}\ , \quad \Upsilon_{S+1}=
\{\Phi_S,H_2\}\ ,
$$
with $\Upsilon_{r}\in \Pc_{r+3}$. Therefore, again by
Lemmas~\ref{lemma:par_poisson}, \ref{lemma:omologico},
\ref{lemma:stima},  we get
$$
\sigma_{\dot \Phi^{(S)}}\le \sigma_{\Upsilon_S}+
\sigma_{\Upsilon_{S+1}} \le \sqrt N C^S_4 (S!)^{5/2} \beta^{-(S+3)/2}\left(1+
\beta^{-1/2}\right) \ .
$$
For  $\beta$ large enough, this estimate and relation
(\ref{eq:minorazione_Phi_s}) give the first statement in
(\ref{eq:ausiliario}) and conclude the proof.

\section{Proof of Theorem~\ref{teor:principale} and
  Corollary~\ref{cor:principale}}\label{sez:corollario}  
The proof of Theorem~\ref{teor:principale} for $E^+$ lays on an application
of Theorem~1 of  
\cite{correlazioni} to the difference
$$
E^+-\Omega \Phi_0= \frac 12\sum_k\left(1-\frac
\Omega{\omega_k^+}\right)\left(\left|\hat p_k^+\right|^2+ \left. \omega^+_k 
\right.^2 \left|\hat q_k^+\right|^2 \right)= \frac 12\sum_k\left(1-\frac
\Omega{\omega_k^+}\right)\xi^+_k\eta^+_k\ .
$$
Indeed, if $m_1/m_2>2$,
$$
\left\|E^+-\Omega\Phi_0\right\|_+=\sup_k\left(1-\frac
\Omega{\omega_k^+}\right) \le \frac{1}{\sqrt 2} \frac{m_2}{m_1}\ , 
$$
so that, on account of Lemma~\ref{lemma:stima}, there exists $C_1>0$
such that
\begin{equation*}
  \begin{split}
&\sigma_{E^+-\Omega\Phi_0}\le \sqrt
N\frac{m_2}{m_1}\frac{C_1}{\beta}\quad\Rightarrow\\
&\sigma_{E^+-\Omega\Phi^{(S)}}\le \sigma_{E^+-\Omega\Phi_0}+ 
\Omega\sigma_{\Phi^{(S)}- \Phi_0}\le \sqrt
N\frac{C_1}{\beta}\left(\frac{m_2}{m_1}+\frac1{\sqrt\beta}\right) \ ,
  \end{split}
\end{equation*}
where, in the second line, use is made of (\ref{eq:ausiliario}).
In a way identical to (\ref{eq:minorazione_Phi_0}) it is then shown
that there exists $C_2>0$ such that
$$
\sigma_{E^+}\ge \sqrt N \frac{C_2}{\beta}\ ,
$$
and thus, by using Theorem~1 in \cite{correlazioni}, there exists
$K_2>0$ such that
$$
\left|C_{E^+}(t)-C_{\Omega\Phi^{(S)}}(t)\right|\le
K_2\left(\frac{m_2}{m_1}+\frac 1{\sqrt\beta}\right) 
\sigma^2_{E^+} \ .
$$
From Theorem~\ref{teor:ausiliario} then Theorem~\ref{teor:principale}
for $E^+$ is 
immediately deduced. 

Coming to the statement for $E^-$, we observe that
$$
E^-=H-E^+- H_{nl}\ ,
$$
where we have defined $H_{nl}=H-H_0$.
Since $H$ is a constant of motion,
$$
\langle H_t\cdot F\rangle= \langle H F_t\rangle= \langle H F\rangle\ ,
$$
for any dynamical variable $F$, thus showing that
\begin{equation}\label{eq:corr_differenza}
  \begin{split}
C_{E^-}(t)-C_{E^-}(0)=& C_{E^+}(t)-C_{E^+}(0)+ C_{H_{nl}}(t)-
C_{H_{nl}}(0)\\
&- \langle
\left(E^+_t-E^+\right) \left(H_{nl}-\langle
H_{nl}\rangle\right)\rangle\\&
- \langle
\left(\left(H_{nl}\right)_t-H_{nl}\right) \left(E^+-\langle E^+\rangle
\right)\rangle \ . 
  \end{split}
\end{equation}
We notice that, because of Lemma~\ref{lemma:stima}, the following
inequalities hold:
$$
\sigma_{H_0-H}\le C_1\frac{\sqrt N}{\beta^{3/2}}\ ,\quad
\sigma_{E^+}\le C_2 \frac{\sqrt N }\beta\ ,
$$
for suitable $C_1,C_2>0$. This, together with the fact that
\begin{equation}\label{eq:corr_nonlineare}
  \begin{split}
    &C_{H_{nl}}(t)\le \sigma^2_{H_{nl}}\ , \\
    &\langle \left(E^+_t-E^+\right) \left(H_{nl}-\langle
    H_{nl}\rangle\right)\rangle \le \sqrt2\sigma_{E^+}\sigma_{H_{nl}}\ ,\\
&    \langle
\left(\left(H_{nl}\right)_t-(H_{nl})\right) \left(E^+-\langle E^+\rangle
\right)\rangle  \le \sqrt2\sigma_{E^+}\sigma_{H_{nl}}\ ,
  \end{split}
  \end{equation}
enable us to infer from (\ref{eq:corr_differenza}) that
$$
\left|C_{E^-}(t)-\sigma^2_{E^-}\right|\le \left|C_{E^+}(t)-C_{E^+}(0)\right|+ 2 C_1^2
\frac{N}{\beta^3} + 2 C_1C_2 \frac{N}{\beta^{5/2}} \ .
$$
Hence, since $\sigma_{E^-}\ge C_3\sqrt N /\beta$ for a suitable
$C_3>0$ and  by the already proved statement for $E^+$, the thesis of
Theorem~\ref{teor:principale} follows.

Corollary~\ref{cor:principale} is then easily deduced, by applying
Cebyshev  inequality to the quantity 
  $$
   \langle
  \left(E^\pm(t)-E^\pm\right)^2\rangle =\frac 12
  \left|C_{E^\pm}(t)-C_{E^\pm}(0)\right|\ .
  $$

\section{Concluding remarks: discussion of the heat capacity of the system}\label{sez:conseguenze}
We have proved that both the energy of the optical branch and the
energy of the acoustic branch are approximately conserved variables,
i.e., their time autocorrelations stay close to the initial value for
long times. This seems to be in contrast with the idea of
thermalization and shows that the system is not mixing on the
considered time scales and exhibits a metastable behaviour. But does
such a lack of ergodicity entail some consequences for the
thermodinamical observables? This is not obvious at all, but in this
particular case we can imagine, in a completely heuristic way, a
mechanism for which this 
slow decay of correlations might show up in the measurement of an
actual physical quantity, the heat capacity $C$ of the chain.
Recall, indeed, that the expression of the heat capacity $C$ of a system put
in contact with a thermostat, in the
linear response theory approximation (see, for instance,
\cite{termomeccanica}), is the following
$$
C(t)=C_H(0)  -C_H(t) \ ,
$$
where $t$ denotes the duration of the measurement process, while the averages
are taken with respect to an invariant measure and the flow is the one
given by the full system (i.e., system of interest with
Hamiltonian $H$, plus thermostat and the interaction terms between the
two). By writing $H=H_0+H_{nl}$ and $H_0=E^++E^-$, such an expression becomes
\begin{equation}\label{eq:calspec_sviluppato}
\begin{split}
C(t) = &\left(C_{E^-}(0)  - C_{E^-}(t)\right)+ 
\left( C_{E^+}(0)  - C_{E^+}(t)\right) \\
&+ \langle \left(E^+_t  -
E^+\right)  \left(E^-_t  - E^-\right)\rangle\\
&+ \left( C_{E^+}(0)  - C_{E^+}(t)\right) +2 \langle
\left(\left(H_0\right)_t -H_0\right)\left(\left(H_{nl}\right)_t
-H_{nl}\right)\ ,
\end{split}
\end{equation}
where the terms in the third line can be neglected for small
temperatures (see formulas (\ref{eq:corr_nonlineare}) above and the subsequent
remarks).

One can imagine  
a thermostat which exchanges energy mainly with one of the two
branches, as it happens if we model the thermostat as a gas of
particles, each interacting with  some of the particles of the FPU
chain through a short--range smooth potential
(see \cite{calspecFPU}). In absence of a mechanism of energy
exchange between the branches, this would imply that, at low
temperatures and for times $t$ of order $\beta^{S/2}$, one has
$$
\langle \left(E^+_t  -
E^+\right)^2\rangle =2 \left( C_{E^+}(0)  - C_{E^+}(t)\right)\ll
\sigma^2_{E^+}\ ,
$$ 
for the complete system, too. Since the second line of
(\ref{eq:calspec_sviluppato}) can be bounded from above by
$$
\langle \left(E^+_t  -
E^+\right)  \left(E^-_t  - E^-\right)\rangle\le 
\sigma_{E^+}\sqrt{2\langle \left(E^+_t  - 
E^+\right)^2\rangle}\ ,
$$
this implies that, for not too long times,
$$
C(t)\approx C_{E^-}(0) - C_{E^-}(t)\ .
$$
As $C_{E^-}(t)$ is expected to decay quickly to zero if the thermostat
is suitably chosen,\footnote{In particular, care should be taken in
  modelling the   thermostat so that the decay rate of this quantity
  does not grow with $N$.} this means
that the measured heat capacity would stabilize around the value
$\sigma^2_{E^-}$, which is significantly smaller than the 
equilibrium value $\sigma^2_H$.

This is, of course, just the rough cast of an idea, but a
result of this kind would be of extreme interest, also in view of the
recent works on metastable behaviour of polymer chains (see
\cite{polimeri,DNA,polimeri2}),
and we plan to work in the near future to establish, following the example of
\cite{vincoli2}, whether a result of this kind can be
proved as a theorem, by suitably choosing the properties of
the thermostat. Alongside this, we plan to tackle the task of
extending the previous result to higher dimensional lattices, in order
to understand whether a similar behaviour could be displayed by real
solids, where the branches of the dispersion relation exhibit a
complex, interlaced structure.

\vskip 2 ex
\noindent
    {\large\textbf{Acknowledgements.}}
    I wish to thank Professors L.~Galgani, A.~Carati and D.~Bambusi for
    their encouragement and for useful comments and discussions.

\appendix
\section{Normal modes of oscillation}\label{app:modi}
We are looking for a change of variables to the normal modes of
oscillation of the form
\begin{equation}\label{eq:trasf_x_q}
\vett x_j=\sum_k \sum_{l=\pm} \vett u^l_k{\hat q}_k^l e^{i\kappa j}\ ,
\end{equation}
i.e., for solutions of the linearized dynamics as
$ \vett x_j=\vett u e^{i(\kappa j-\omega t)}$, with $\kappa=2\pi k/N$, and
$k=\lfloor-N/2\rfloor+1, \ldots, \lfloor N/2\rfloor$.

Corresponding to the frequencies
$$
\left(\omega_k^\pm\right)^2= K\frac{m_1+m_2\pm \sqrt
  \Delta_k}{m_1m_2}\ , \quad\mbox{in which }
\Delta_k=m_1^2+m_2^2+2m_1m_2 \cos\frac{2\pi k}{N}\ ,
$$
a solution for $\vett u^\pm_k$ is
\begin{equation*}
\begin{split}
&\vett u^\pm_k=c^\pm_k\left(\begin{array}{c} \cos \tfrac \kappa2\\
  (m_2-m_1\mp \sqrt
  \Delta_k)e^{i\kappa/2}/2m_2 \end{array}\right)\ ,\quad \mbox{for }
k\neq N/2\ ,\\
&\quad \vett u^+_{N/2}=\frac1{\sqrt N}\left(\begin{array}{c} 0\\ 
  1/\sqrt{m_2}\end{array}\right)\ , \quad \vett
u^-_{N/2}=\frac1{\sqrt N}\left(\begin{array}{c} 1/\sqrt{m_1}\\  
  0 \end{array}\right)\ ,
\end{split}
\end{equation*}
where the second line is needed only if $N$ is even and we have
introduced a normalization factor
$$
c^\pm_k=\left(\frac{N\sqrt{\Delta_k}}{2m_2} \left(\sqrt{\Delta_k}\mp 
(m_2-m_1)\right)\right)^{-1/2}\ .
$$
Such a normalization is so chosen that, for the Hermitian product in
$\mathbb C^2$, denoted by $\langle \cdot,\cdot\rangle$, it holds
\begin{equation}\label{eq:normalizzazione_u}
\langle\vett u_k^l, M\vett u_{k'}^{l'}\rangle= \frac 1N \delta_{k,k'}
\delta_{l,l'}  \ ,\qquad \mbox{with } M=\left(\begin{array}{cc}
  m_1& 0\\
  0& m_2\end{array} \right)\ .
\end{equation}
Notice that, since $\vett x_j$ are real, while $\vett
u^\pm_k=\bar{\vett u}^\pm_{-k}$, the complex coordinates (coordinates
of the normal modes) $\hat q^\pm_k$ satisfy the relations
$\hat q^\pm_{-k}=\bar{\hat q}^\pm_k$, for $k\neq 0,N/2$, whereas they
are real for $k=0,N/2$.

In order to invert (\ref{eq:trasf_x_q}) we start from relation
\begin{equation}\label{eq:trasf_q_x}
\vett u_k^+\hat q_k^++\vett u_k^-\hat q_k^-=\sum_j \vett x_j e^{-i\kappa j}\ ,
\end{equation}
and take the Hermitian product, respectively, with $M\vett u_k^+$ and
$M\vett u_k^-$. We thus get
\begin{equation}\label{eq:trasf_q_x_fin}
  \begin{split}
\hat q_k^\pm&=\sum_j \langle\vett x_j,M\vett u_k^\pm \rangle e^{-i\kappa
  j}=\sum_j\left(m_1 x_{j,1} \re u_{k,1}^\pm +
m_2 x_{j,2} \re u_{k,2}^\pm\right)  e^{-i\kappa
  j} \ .
\end{split}
\end{equation}
For the conjugate moments $\hat p_k^\pm$, the condition of canonicity
imposes then 
\begin{equation}\label{eq:trasf_p_p}
\hat p_k^\pm=\sum_j \left(\frac{p_{j,1}}{\sqrt{m_1}} \re\sqrt m_1 u_{k,1}^\pm +
\frac{p_{j,2}}{\sqrt{m_2}} \re\sqrt{m_2} u_{k,2}^\pm\right)  e^{i\kappa
  j}\ .
\end{equation}
Remark that, on account of (\ref{eq:normalizzazione_u}), $|\re\sqrt
{m_i} u_{k,i}|\le 1/\sqrt N$.  

\subsection{The transformation to the difference coordinates}
In order to express the Hamiltonian as a function of the normal modes
coordinates, it will be useful to write explicitely the relation
between them and  the difference coordinates $\vett
r_j\equal(x_{j,2}-x_{j,1},x_{j,1}-x_{j-1,2})$, i.e.,
\begin{equation}\label{eq:trasf_r_q}
\vett r_j=\sum_k \left( \vett w_k^+{\hat q}_k^++ \vett w_k^-{\hat q}_k^-\right)
e^{i\kappa j}\ , \quad\mbox{with } \vett w_k^\pm=\left(\begin{array}{c}
    u^\pm_{k,2}-u^\pm_{k,1}\\
    u^\pm_{k,1}-u^\pm_{k,2}e^{-i\kappa} \end{array}\right)
\end{equation}
Here, an explicit calculation shows that
$$
\vett w_k^\pm= c_k^\pm\left(
\begin{array}{c}
-\frac{m_1\left.\omega_k^\pm\right.^2 \cos \tfrac \kappa 2}{K} +
i\tfrac{m_2-m_1\mp\sqrt{\Delta_k}}{2m_2} \sin \tfrac \kappa 2\\
\frac{m_1\left.\omega_k^\pm\right.^2\cos \tfrac \kappa 2}K +
i\tfrac{m_2-m_1\mp\sqrt{\Delta_k}}{2m_2} \sin \tfrac \kappa2
\end{array}
\right)= \frac{\omega_k^\pm}{\sqrt{2NK}} \left(
\begin{array}{c}
e^{i\alpha_k^\pm}
  \\
-e^{-i\alpha_k^\pm}
\end{array}
\right)\ ,
$$
with the complex phase $\alpha_k$ determined by
$$
e^{2i\alpha_k^\pm}=\pm\frac{m_2+m_1e^{i\kappa}}{\sqrt \Delta_k}\ .
$$
From here, it can be immediately shown that, for any $m$,
\begin{equation}\label{eq:potenziale_modi}
  \begin{split}
\sum_{j=1}^Nr_{j,m}^s&=\sum_{k_1,\ldots,k_s}\sum_{l_1,\ldots,l_s=\pm}
\left( w_{k_1,m}^{l_1}{\hat q}_{k_1}^{l_1}\cdots w_{k_s,m}^{l_s}{\hat q}_{k_s}^{l_s}\right) \sum_{j=1}^Ne^{i(\kappa_1+\cdots+\kappa_s)j} \\
&=\sum_{k_1,\ldots,k_s}\sum_{l_1,\ldots,l_s=\pm} \left( w_{k_1,m}^{l_1}{\hat q}_{k_1}^{l_1}\cdots
w_{k_n,m}^{l_n}{\hat q}_{k_n}^{l_n}\right) \sum_n\delta_{k_1+\cdots+k_s}^n\ ,
  \end{split}
\end{equation}
where we made use of
\begin{equation*}
\sum_{j=1}^N e^{i2\pi k j/N}=N \sum_{n \in \mathbb Z} \delta_k^n\ , \quad \mbox{with }
\delta_{k}^n=\delta_{k,nN}\ ,
\end{equation*}
which is valid for any integer $k$. This is particularly relevant,
because  the perturbing parts of the Hamiltonian can be written as
$$
H_1=\frac{KA}2 \sum_j\left(r_{j,1}^3+r_{j,2}^3\right)\ ,\quad
H_2=\frac{KB}2 \sum_j\left(r_{j,1}^4+r_{j,2}^4\right)\ ,
$$
so that it is immediately seen that they belong to $P_3$ and $\mathcal
P_4$, respectively.

Relation (\ref{eq:trasf_r_q}) can then be easily
inverted, by using Fourier series properties, which give
\begin{equation}\label{eq:trasf_q_r_1}
\vett w_k^+
{\hat q}_k^+ + \vett w_k^- {\hat q}_k^-
=\frac 1N \sum_{j=1}^N \vett r_j e^{-i\kappa j}\ .
\end{equation}
We are however interested in equations which rely separately
${\hat q}_k^+$ and ${\hat q}_k^-$ to $\vett r_j$. Since $\vett w_k^+$
and $\vett w_k^-$ are orthogonal with respect to 
the Hermitian product in $\mathbb C^2$ and $\langle
\vett w_k^\pm,\vett w_k^\pm\rangle= \left.\omega_k^\pm\right.^2/(NK)$, we multiply both
sides of (\ref{eq:trasf_q_r_1}) by
 $\vett w_k^+$ and $\vett w_k^-$, and get
$$
\frac{\left.\omega_k^\pm\right.^2{\hat q}_k^\pm} {NK}= \frac 1N\sum_j
\langle \vett r_j,\vett 
w_k^\pm\rangle e^{-i\kappa j}= \frac {\omega_k^\pm }{N\sqrt{ 2 
  NK}}\sum_j \left(r_{j,1} 
\cos \alpha_k^\pm -r_{j,2}\cos\alpha_k^\pm
\right)e^{-i\kappa j}\ .
$$
From here, the crucial equality follows
\begin{equation}\label{eq:trasf_q_r}
\omega_k^\pm{\hat q}_k^\pm= \sqrt{\frac{K}{2N}}\sum_j \left(r_{j,1}
\cos\alpha_k^\pm -r_{j,2}\cos\alpha_k^\pm
\right)e^{-i\kappa j}\ .
    \end{equation}


\section{Proof of Lemma~\ref{lemma:par_poisson}}\label{app:par_poisson}
We can  write explicitly the Poisson brackets as
\begin{equation*}
\begin{split}
\{f,g\}=& \frac{i}{N^{(r+s-4)/2}}\sum_{(\sigma,\tilde \tau,k,l,n)\in
  \I_s}\sum_{(\sigma',\tilde \tau',k',l',n')\in \I_r}
f_{\sigma,\tilde \tau,l,n}\,g_{\sigma',\tilde \tau',l',n'}\\
& \sum_{j=1}^s\sum_{m=1}^r
\sigma_j
\frac{ \omega^{l_j}_{k_j}  \Xi^s_{\sigma,k,l}\Xi^r_{\sigma',k',l'}
}{\xi^{l_j}_{k_j}
  \eta_{k_j}^{l_j}}\delta_{\sigma_j,-\sigma'_m}\delta_{k_j,k'_m}\delta_{l_j,l'_m}
\delta^n_{\tilde \tau\cdot k} \delta^{n'}_{\tilde \tau'\cdot k'}\ .  
\end{split}
\end{equation*}
We exchange the order of the sums over $j$ and $m$ with those over
$(\sigma,\tilde \tau,k,l,n)$ and $(\sigma',\tilde \tau',k',l',n)$, by
summing first over  $(\sigma_j,k_j,l_j,\tilde \tau,n)$ and
$(\sigma'_m,k'_m,l'_m,\tilde \tau',n')$: this gives
\begin{equation*}
\begin{split}
\{f,g\}=& \frac{i}{N^{(r+s-4)/2}}\sum_{j=1}^s\sum_{m=1}^r
\sum_{\indice,\indice'}\Xi^{s-1}_{\indice}\Xi^{r-1}_{\indice'}\sum_{\sigma_j,k_j,l_j,\tilde
  \tau, n}\\
&\sum_{\sigma'_m,k'_m,l'_m,\tilde \tau', n'}
f_{\sigma,\tilde \tau,l,n}\,g_{\sigma',\tilde \tau',l',n'}
 \sigma_j \omega^{l_j}_{k_j}  
\delta_{\sigma_j,-\sigma'_m}\delta_{k_j,k'_m}\delta_{l_j,l'_m}\delta^n_{\tilde
  \tau\cdot k} \delta^{n'}_{\tilde \tau'\cdot k'} 
\end{split}
\end{equation*}
where
\begin{equation*}
\begin{split}
\mathfrak i=(\{\sigma_i\}_{i\neq
  j},\{k_i\}_{i\neq j},\{l_i\}_{i\neq j})\ ,\quad
\mathfrak i'=(\{\sigma'_i\}_{i\neq
  m},\{k'_i\}_{i\neq m},\{l'_i\}_{i\neq
  m})\ .
\end{split}
\end{equation*}
We note that
$$
\Xi^{s-1}_{\mathfrak
  i}\Xi^{r-1}_{\mathfrak i'}= \Xi^{r+s-2}_{\sigma'',k'',l''}\ ,
\quad \mbox{with }
 \sigma''= \{\sigma_i\}_{i\neq
  j}\cup\{\sigma'_i\}_{i\neq m}
\ ,\ldots
$$
is a monomial of degree $r+s-2$. We can therefore write $\{f,g\}=h\in
\Pc_{r+s-2}$, with
\begin{equation*}
\begin{split}
h=& \frac{1}{N^{(r+s-4)/2}}\sum_{(\sigma'',\tilde
  \tau'',k'',l'',n'')\in \I_{r+s-2} }  h_{\sigma'',\tilde \tau'',
  l'',n''}\Xi^{r+s-2}_{\sigma'',k'',l''}\delta^{n''}_{\tilde \tau''\cdot
  k''}\ .
\end{split}
\end{equation*}
Here
\begin{equation*}
\begin{split}
h_{\sigma'',\tilde \tau'',  l'',n''}\delta^{n''}_{\tilde \tau''\cdot
  k''} = &i \sum_{j=1}^s\sum_{m=1}^r\sum_{\sigma_j,k_j,l_j,\tilde
  \tau_j, n}\sum_{\sigma'_m,k'_m,l'_m,\tilde \tau'_m, n'}
 \sigma_j \omega^{l_j}_{k_j} \\ 
&f_{\sigma,\tilde \tau,l,n}\,g_{\sigma',\tilde
  \tau',l',n'}\delta_{\sigma_j,-\sigma'_m}\delta_{k_j,k'_m}\delta_{l_j,l'_m}\delta^n_{\tilde 
  \tau\cdot k} \delta^{n'}_{\tilde \tau'\cdot k'}\delta_{n'',\tilde \tau_jn-\tilde\tau'_m n'}\ , 
\end{split}
\end{equation*}
where $(\sigma,\tilde \tau,k,l)$ and $(\sigma',\tilde \tau',k',l')$
are determined in terms of $(\sigma'',\tilde \tau'',k'',l'')$,
$\sigma_j,k_j,l_j,\tilde 
\tau_j,\sigma'_m,k'_m,l'_m,\tilde \tau'_m$. Indeed, one has
\begin{equation*}
  \begin{split}
  \tilde \tau&= \{\tilde \tau_j\tilde
  \tau''_i\}_{i<j}\cup\{\tilde\tau_j\}\cup\{\tilde\tau_j\tilde\tau''_i\}_{j\le
    i<s} \ ,\\ 
\tilde\tau'&=\{-\tilde\tau'_m\tilde \tau''_i\}_{s\le
  i<s+m-1}\cup\{\tilde\tau'_m\}\cup\{-\tilde\tau'_m\tilde\tau''_i\}_{i\ge
  s+m-1}\ ,\\ 
\sigma&=\{\sigma''_i\}_{i<j}\cup\{\sigma_j\}\cup\{\sigma''_i\}_{j\le i<s}\ , \\
\sigma'&=\{\sigma''_i\}_{s\le
  i<s+m-1}\cup\{\sigma'_m\}\cup\{\sigma''_i\}_{i\ge s+m-1}\ ,  
  \end{split}
\end{equation*}
and similar relations for $k,k',l,l'$. Due to the appearance of
Kronecker deltas, $\sigma'_m=-\sigma_j$, 
$l'_m=l_j$ and $k'_m=k_j$, whereas $ \delta^n_{\tilde 
  \tau\cdot k}$ imposes $k_j=\tilde \tau_jn-\sum_{i\neq j}
  \tilde\tau_j\tilde\tau_i $ and $\delta_{n'',\tilde
    \tau_jn-\tilde\tau'_m n'}$ fixes $n'=\tilde \tau'_m \tilde \tau_j
  n-\tilde \tau'_m n''$. Such remarks enable us
  to estimate the norm of $h$ by summing only on the free indices as
$$
\left\|h\right\|_+\le \Omega\left\|f\right\|_+\left\|g\right\|_+
\sum_{j=1}^s\sum_{m=1}^r \sum_{\sigma_j,\tilde \tau_j,\tilde \tau'_m=\pm
  1} \sum_{l_j=\pm}\sum_{n=-\lfloor (s-1)/2\rfloor}^{\lfloor
(s-1)/2\rfloor} 1\ .
$$ 
By possibly exchanging the
role of  $n$ with 
that of $n'$, the thesis is got.

\section{Proof of Lemma~\ref{lemma:omologico}}\label{app:omologico}
 For what concerns the operator norm of $K$, we note that, if $g\in
 \resto_s$, $\|L_\Omega g\|_+\ge \Omega 
\|g\|_+$.  Since on $\Pc_s$
$$
\|L_{\Theta_0}\|\le s \max\{\max \omega_k^+-\min \omega_k^+,\max
\omega_k^-\}= \sqrt{\frac{2K}{m_1}}s\le s \Omega\sqrt{\frac{m_2}{m_1}} \ ,
$$
the norm of $K$  on
$\resto_s$ is smaller than 1/2 for $s\le S$. Thus, for any $g\in
\resto_s$ it holds $\|L_0^{-1}f\|_+\le 2\|f\|_+/\Omega$, by Neumann
inversion formula. 

Coming to the solutions of the homological equation (which are
nontrivial only for $S\ge 3$), we observe first of all that
$\Psi_1=H_1\in \Pc_3$, so that $Z_1$ can be chosen as 
the projection over $\N_1$ of $\Psi_1$. This implies that
$Z_1-\Psi_1\in \resto_3$, and eq.~\eqref{eq:omologica} can be
accordingly solved,  with $\chi_1\in \Pc_3$, and $\|\chi_1\|_+\le
2\|\Psi_1\|_+/\Omega$. The estimate  
$\|Z_1\|_+\le\|\Psi_1\|_+= \|H_1\|_+\le B$ is then valid, in
agreement with (\ref{eq:induzione1}) for $s=1$. For $\Psi_2$, in a
similar way, it can be seen that it belongs to $\Pc_4$, so as $Z_2$
and $\chi_2$, while its norm,  due to Lemma~\ref{lemma:par_poisson},
is bounded from above by 
$$
\left\|\Psi_2\right\|_+\le \left\|H_2\right\|_++
\frac12\left(\left\|L_{\chi_1}H_1 \right\|_++\left\|L_{\chi_1}Z_1
\right\|_+\right) \le B^2(2+2^5\cdot 3^3)\ , 
$$
which satisfies (\ref{eq:induzione1}) if $C\ge 2^63^3$.

For all other orders we proceed by induction, observing that, for
$s\le S$, $\Psi_s\in \Pc_{s+2}$ on account of
Lemma~\ref{lemma:par_poisson}, and, as a consequence $Z_s$ and
$\chi_s$ belong to $\Pc_{s+2}$, too. This entails, as previously observed,
that $\|\chi_s\|_+\le 2\|\Psi_s\|_+/\Omega$, for $s\le r-1$. Since the
expression for  $\Psi_r$ involves $E_sZ_l$, for $1\le s\le r-1$, and
$1\le l\le r-1$,  let us suppose that hypothesis
(\ref{eq:induzione1}) is true for $s\le r-1$ and prove first formula
(\ref{eq:induzione2}) by induction on $s$, for $1\le s\le r-1$ and
$l\ge 1$ fixed. For $s=1$ this is trivially done  by using the
fact that $n+1\le 2n$ if $n\ge1$, as
\begin{equation*}
  \begin{split}
\left \|E_1f_l\right\|_+&=\left \|L_{\chi_1}f_l\right\|_+\le 2^53^2B 
 (l+2)\left \|f_l\right\|_+  \\
&\le \frac 14BC \frac{(l+1)!}{l!}\left \|f_l\right\|_+\ ,
  \end{split}
\end{equation*}
if $C\ge 2^83^2$. For $s=2$, instead, we write
\begin{equation*}
  \begin{split}
\left \|E_2f_l\right\|_+&=\left \|L_{\chi_2}f_l\right\|_+ + \frac12
\left \|L^2_{\chi_1}f_l\right\|_+\\
&\le 2^9B^2\left \|f_l\right\|_+\left(2C
(l+2)+3^4(l+2)(l+3)\right)  \\
&\le (BC)^2 \left \|f_l\right\|_+\left(\frac{2^{10}}{C}\frac{(l+2)!}{(l+1)!}
+ \frac{2^93^4}{C^2}\frac{(l+2)!}{l!}\right) \ .
  \end{split}
\end{equation*}
This is in agreement with (\ref{eq:induzione2}) if $C\ge
2^{12}$. For what concerns the terms with $ 3\le s\le r-1$, we
repeatedly use the elementary inequality
\begin{equation}\label{eq:binomiali}
  n!m!\le k!(n+m-k)!\ ,  \quad\mbox{if }n,m\ge k\ ,
  \ .
\end{equation}
in the following chain of inequalities
\begin{equation*}
  \begin{split}
\left\|E_sf_l\right\|_+&\le\left\|L_{\chi_s}f_l\right\|_++\left\|L_{\chi_{s-1}}
 L_{\chi_1}  f_l\right\|_+ +\frac 1s
 \left\|L_{\chi_1}E_{s-1} f_l\right\|_+\\
 &+\frac 2s
\left\|L_{\chi_2}E_{s-2} f_l\right\|_++\sum_{j=3}^{s-2} \frac js
\left\|L_{\chi_j}E_{s-j}f_l\right\|_+\\ 
&\le B^sC^{s-1}\left(2^6+\frac{2^{11}3^2}{C}\right)
(s+1)! \frac{(l+3)!}{(l+1)!}\left\|f_l\right\|_+ \\
&+2^6 3^2B\frac{s+l}s\left\|E_{s-1} f_l\right\|_+ 
+2^{12}B^2C\frac{s+l-1}{s}\left\|E_{s-2} f_l\right\|_+ \\
&+\frac
{2^5}{s}\sum_{j=3}^{s-2}B^jC^{j-1}(j+2)!(s-j+l+2)^2\left\|E_{s-j} f_l\right\|_+ \\
&\le 2^63B^sC^{s-1}\frac{(s+l)!}{l!} \left\|f_l\right\|_+
\Biggl(\left(2^3+\frac{2^83^2}{C}\right)\frac{1}{l+1} \\
&+
\frac{3}{4}\left(\frac{1}{s!} +\frac
1{s(l+1)}\right) 
+\frac{2^4}{l+1}+ \frac2{l+1} \Biggr)\ , 
  \end{split}
\end{equation*}
which satisfies (\ref{eq:induzione2}) if $C\ge 2^{13}$.

We come then to the other inductive hypothesis, i.e.,
(\ref{eq:induzione1}), noticing that, by the very definition of
$\Psi$,  one has
$$
\left\|\Psi_r\right\|_+\le \left\|H_r\right\|_+ + \sum_{j=1}^{r-1}
\frac{1}{r} \left(j\left\|L_{\chi_j}H_{r-j}\right\|_++ j\left\|E_{r-j}
Z_j\right\|_+ \right)\ .
$$
We treat separately the single terms, making use of inequalities
(\ref{eq:binomiali}). The first addendum is trivially bound from above
by hypothesis, while the first term in brackets is bounded from above,
due to Lemma~\ref{lemma:par_poisson}, via
\begin{equation*}
  \begin{split}
j\left\|L_{\chi_j}H_{r-j}\right\|_+&\le 2^5j\left\|\Psi_j \right\|_+
\left\|H_{r-j} \right\|_+ (j+2)(r-j+2)\min(j+2,r-j+2)\\
&\le 2^6B^rC^{j-1} (j+2)! (r-j+2)! \\
&\le 2^93B^rC^{j-1} r!\ ,\quad\mbox{ for } 2\le j\le r-2\ ,\\
\left\|L_{\chi_1}H_{r-1}\right\|_+&\le 2^63^2B^r  r!\ , \quad\quad
\left\|L_{\chi_{r-1}}H_1\right\|_+\le 2^63^2B^rC^{r-2}  r!\ . \quad 
  \end{split}
\end{equation*}
The other term in brackets, by  inductive hypotheses
(\ref{eq:induzione1}--\ref{eq:induzione2}) at the previous orders, is
bounded from above by 
$$
\sum_{j=1}^{r-1}
\frac{j}{r}\left\|E_{r-j}Z_j\right\|_+\le \frac{B^rC^{r-1}}{4} r!
\sum_{j=1}^{r-1} \left(\frac{1}{(r-j)!}+\frac{j}{r(j+1)}\right)
\le\frac{B^rC^{r-1}}{4} r!(e+1) \ .
$$
Taking the sum over $j$, this gives
$$
\left\|\Psi_r\right\|_+\le B^rr!\left(1+ 2^93C^{r-3}+2^63^2C^{r-2}+ \frac{e+1}4
C^{r-1}\right) \ ,
$$
from which (\ref{eq:induzione1}) follows, for $C\ge 2^{10}3^2$.

Only the proof of (\ref{eq:induzione2}) for $l=0$ is left, as it was
not needed above. This is
trivially true for $s=1$, while for $2\le s\le S$, by induction one has
\begin{equation*}
  \begin{split}
\left\|E_sf_l\right\|_+&\le\left\|L_{\chi_s}f_l\right\|_+ +
\left\|L_{\chi_{s-1}}L_{\chi_1}f_l\right\|_++\frac
1s\left\|L_{\chi_1}E_{s-1}f_l\right\|_+ +\sum_{j=2}^{s-2} \frac js  
\left\|L_{\chi_j}E_{s-j}f_l\right\|_+\\ 
&\le
2^5B^sC^{s-1}\left\|f_l\right\|_+\Biggl(\left(2^3+ \frac{2^83^2}{C}+2\cdot3\right)
(s+1)!\\  &\quad\quad +\frac
1{4s}\sum_{j=2}^{s-2}(j+2)!(s-j+3)! \Biggr) \\
&\le \left(2^{10}+\frac{2^{13}3^3}{C}\right)B^sC^{s-1}\left\|f_l\right\|_+(s+1)!\ ,
  \end{split}
\end{equation*}
whence (\ref{eq:induzione2}) for $C\ge 2^{12}$.

\section{Proof of Lemmas~\ref{lemma:somma_scambiabili} and \ref{lemma:indipendenti}}\label{app:scambiabili}
The proof of both Lemmas is performed by expressing the monomials
$\Xi^s_{\sigma,k,l}$ in the coordinates $(\vett p_j,\vett r_j)$, in
which the Gibbs measure is easier to control. In fact, it possesses
several remarkable properties:
\begin{itemize}
  \item the variables $\vett p_j$ and $\vett r_{j'}$ are mutually
    independent;
  \item the variables $r_{j,i}$ are exchangeable (see
    \cite{scambiabili} for the concept of exchangeability);
  \item the variables  $p_{j,i}$ are pairwise independent (and so, in
    particular, they are exchangeable).
\end{itemize}
In addition, we have an estimate of the mean values of the monomials
in such variables, which is expressed in subsequent
Lemma~\ref{lemma:misura}. There, we denote by
$$\vett
y_j=(p_{j,1}/\sqrt{m_1},p_{j,2}/\sqrt{m_2},\sqrt K r_{j,1},\sqrt 
K r_{j,2})\ ,
$$
and by
\begin{equation*}
  \begin{split}
&y^s_{j,\alpha}=y_{j_1,\alpha_1}\cdots y_{j_s,\alpha_s}\ ,\quad\mbox{for }
j=(j_1,\ldots,j_s)\ ,\\
& \alpha=(\alpha_1,\ldots,\alpha_s)\ ,\quad \alpha_i=1,\ldots,4\ ,
  \end{split}
\end{equation*}
while $J$ denotes the vector $J=(j,j')$, i.e., a vector of $s+s'$
components if $j$ and $j'$ have, respectively, $s$, and $s'$
components, which has as first $s$ components those of $j$, then those
of $j'$.
\begin{lemma}\label{lemma:misura}
Let
$\tau=(\tau^{(1)},\ldots,\tau^{(S_1)})$
be a  $s+s'$--admissible collection of indices and let
$\mathcal J(\tau)$ be the set of vectors $J$ such that
\begin{equation}
  \label{selection}
  J_l= J_{l'} \quad \Longleftrightarrow \quad\exists\, i\ \mbox{s.t. }
  l,l'\in\supp(\tau^{(i)}) \ . 
\end{equation}
  Then there exist $K,N_0>0$ such that, for any $s$ and $s'$, any
  $\alpha,\alpha'$, any $\tau$, and any $J\in \mathcal J(\tau)$, one
  has for $N>N_0$ 
\begin{equation}\label{eq:varianza_stessi_siti}
\left|\langle y^s_{j,\alpha} y^{s'}_{j',\alpha'}\rangle -\langle y^s_{j,\alpha}\rangle \langle y^{s'}_{j',\alpha'}\rangle\right| \le
 K^{s+s'}\beta^{-(s+s')/2} \prod_{i=1}^{S_1}\sqrt{\mathfrak n_i!}   \ ,
\end{equation}
in which $\mathfrak n_i$ denotes the cardinality of  $\supp(\tau^{(i)})$.
Moreover, if $(\tau^{(1)},\ldots,\tau^{(S_1)})\in \bar
\Tc$ then
\begin{equation}\label{eq:siti_diversi}
\left|\langle y^s_{j,\alpha} y^{s'}_{j',\alpha'}\rangle -\langle y^s_{j,\alpha}\rangle
\langle y^{s'}_{j',\alpha'} \rangle\right| \le
\frac 1N  K^{s+s'} \beta^{-(s+s')/2} \prod_{i=1}^{S_1}\sqrt{\mathfrak n_i!}\ .
\end{equation}
\end{lemma}

This lemma is a minor modification of Lemma~4 of \cite{primopasso} and
consists in a simple adaptation of standard probabilistic arguments,
which are not reported here.

We pass from the variables $(\xi^\pm,\eta^\pm)$ to $(p^\pm,q^\pm)$ by
using (\ref{eq:trasf_xi_q}), then apply
(\ref{eq:trasf_p_p},\ref{eq:trasf_q_r}) of Appendix~\ref{app:modi} to
pass to the variables $(\vett p_j,\vett r_j)$, thus getting
$$\Xi^s_{\sigma,k,l}=
\sum_{\alpha,\tilde \tau}\frac{c_{k,\alpha,\sigma,l,\tilde \tau}}{2^{s/2}N^{s/2}}\sum_{j_1,\ldots,j_s=1}^N
y^s_{j,\alpha}e^{i\tilde \tau_1\kappa_1 j_1}\cdots e^{i\tilde \tau_s\kappa_s j_s}\ , \quad\mbox{with
} |c_{k,\alpha,\sigma,l,\tau}| \le 1\ ,
$$
where $\tilde \tau=(\tilde\tau_1,\ldots,\tilde \tau_s)$, $\tilde
\tau_i=\pm 1$. Hence follows that
 \begin{equation}\label{eq:correlazione_y}
    \begin{split}
\left| \langle
\Xi^s_{\sigma,k,l}\right.& \left.\Xi^{s'}_{\sigma',k',l'}\rangle - \langle 
\Xi^s_{\sigma,k,l}\rangle \langle\Xi^{s'}_{\sigma',k',l'}\rangle
\right| \le
\frac{1}{2^{(s+s')/2}N^{(s+s')/2}}\sum_{\alpha,\alpha'}\sum_{\tilde \tau,\tilde \tau'}\\
&\left|\sum_{j_1,\ldots,j_s,j'_1,\ldots,j'_{s'}}e^{i\tilde\tau_1\kappa_1
  j_1}\cdots e^{i\tilde\tau'_s\kappa'_{s'} j'_{s'}} 
\left( \langle  y^s_{j,\alpha} y^{s'}_{j',\alpha'}\rangle -\langle y^s_{j,\alpha}\rangle
\langle y^{s'}_{j',\alpha'}\rangle\right)\right| \ .
    \end{split}
 \end{equation}
The sum in the second line has $N^{s+s'}$ terms, which, however, contain
the oscillating factors $e^{i\kappa j}$. The key remark here is that
the property of exchangeability of the variables $y_{j,\alpha}$
entails that the terms in brackets take always the same value, but for
some very peculiar cases, so that almost all oscillating sums
vanish. In fact, let us consider the sequences of complex numbers
$B_J$, with $J=(j_1,\ldots,j_r)$, having the following property:
\begin{definizione}[Property A] Let $\{i_1,\ldots,i_r\}$ be a
  permutation of $\{1,\ldots,r\}$ and let the values of the indices
  $j_{i_1},\ldots,j_{i_n}$ be fixed, for $n<r$, while the remaining
  indices have the same value $j_{i_{n+1}}= \ldots=j_{i_r}=\bar
  \jmath$. We say that the sequence $B_J$ possesses Property A if and
  only if it takes the same value for all values of $\bar \jmath\neq
  j_{i_l}$, for any $l\le n$.
\end{definizione}
Because of exchangeability $\langle
y^s_{j,\alpha} y^{s'}_{j',\alpha'}\rangle -\langle y^s_{j,\alpha}\rangle 
\langle y^{s'}_{j',\alpha'}\rangle$ has precisely this property, for
$J=(j,j')$,  $r=s+s'$ and any $\alpha,\alpha'$. For
this reason, in its estimate we can use the following
\begin{lemma}\label{lemma:somme_su_j}
Let $B_J$ be a sequence satisfying Property $A$ and let
$$
\hat B^{\tilde \tau}_K=\sum_J B_J e^{i\tilde \tau_1\kappa_1j_1}
\cdots e^{i\tilde \tau_r\kappa_rj_r}\ .  
$$
Then
\begin{equation}\label{eq:btau}
\begin{split}
\hat B^{\tilde
  \tau}_K=&\sum_{S_1=1}^rN^{S_1}\sum_{(\tau^{(1)},\ldots,\tau^{(S_1)})\in
\Tc_r}\sum_{n_1,\ldots,n_{S_1}} \delta^{n_1}_{\tau^{(1)}\cdot
    K}\cdots \delta^{n_{S_1}}_{\tau^{(S_1)}\cdot K}\\
&  \sum_{i=1}^{S_1}\sum_{S_2(i)=1}^{\card_i}\sum_{\buildrel{\tau^{(i,1)}+\cdots+\tau^{(i,S_2(i))}
    =\tau^{(i)}} \over {(\tau^{(i,1)},\ldots,\tau^{(i,S_2(i))})\in
    \Tc_{\card_i}(\tau^{(i)})}}  c_{\tilde \tau}^{\tau,S_1}\
\end{split}
\end{equation}
where $\card_i$ is the cardinality of $\supp(\tau^{(i)})$,
$\Tc_{\card_i}(\tau^{(i)})$ is the set of vectors in $\Z_3^r$ which are 
$\card_i$--admissible on the support of  $\tau^{(i)}$ and vanishing
outside it, while
$$
{\tau}=(\tau^{(1,1)},\ldots,\tau^{(1,S_2(1))},
\ldots,\tau^{(S_1,1)}, \ldots,\tau^{(S_1,S_2(S_1))})\in \Tc_r\ .
$$
For the constants $c_{\tilde \tau}^{\tau,S_1}>0$ it holds
\begin{equation}\label{eq:btau_stima}
c_{\tilde \tau}^{\tau,S_1}\le \sup_{J\in \mathcal
  J(\tau)}\left| B_J\right|\prod_{i=1}^{S_1} (S_2(i)-1)!
\ ,
\end{equation}
in which
$\mathcal J(\tau)$ is the set of vectors $J$ defined by (\ref{selection}).

\end{lemma}

The proof of the previous lemma is performed by summing over all $j$,
from $j_r$ down to $j_1$, and observing by induction on $0\le R< r$
that the terms obtained by summing over $j_r,\ldots, j_{r-R}$ have a
peculiar form which we detail now. First of all, let us denote by
$\Tc_r^R$ the set of $(\tau^{(1)},\ldots,\tau^{(r-R+S_1)})\in \Tc_r$
such that $\supp(\tau^{(i)}) \cap \{1,\ldots,r-R\}=i$ for $i\le
r-R$ and with $\Tc^R_{\card_i}(\tau^{(i)})$ the analogous of
$\Tc_{\card_i}(\tau^{(i)})$, with $\Tc_r$ replaced by $\Tc_r^R$; let us put
$$
\tau_R\equal (\tau^{(1,1)},\ldots,\tau^{(1,S_2(1))},\ldots, \tau^{(r-R+S_1,1)},
\ldots, \tau^{(r-R+S_1,S_2(r-R+S_1))})\ .
$$
The inductive hypothesis is the following
\begin{equation}\label{eq:induzione_somme_su_j}
  \begin{split}
\hat B^{\tilde
  \tau}_K =& \sum_{(\tau^{(1)},\ldots,\tau^{(r-R+S_1)})\in
  \Tc_r^R} \sum_{j_1,\ldots, j_{r-R}}e^{i2\pi j_1 K\cdot
  \tau^{(1)}/N}\cdots e^{i2\pi j_{r-R} K\cdot
  \tau^{(r-R)}/N}\\
&\sum_{S_1=0}^RN^{R}\sum_{n_1,\ldots,n_{S_1}} \delta^{n_1}_{\tau^{(r-R+1)}\cdot 
    K}\cdots \delta^{n_{S_1}}_{\tau^{(r-R+S_1)}
  \cdot K}\\
&\sum_{i=1}^{r-R+S_1}\sum_{S_2(i)=1}^{\card_i}\sum_{\buildrel{\tau^{(i,1)}+\cdots+\tau^{(i,S_2(i))}
    =\tau^{(i)}} \over {(\tau^{(i,1)},\ldots,\tau^{(i,S_2(i))})\in
    \Tc^R_{\card_i}(\tau^{(i)})}}
B^{\tau_R,S_1}_{j_1,\ldots,j_{r-R}}\ ,
  \end{split}
\end{equation}
in which the coefficients $B^{\tau_R,S_1}_{j_1,\ldots,j_{r-R}}$ have
Property A with respect to the set of indices $(j_1,\ldots,j_{r-R})$.

By the very definition of $\hat B^{\tilde \tau}_K$ this is true for $R=0$,
putting $\tau^{(i)}_j=\delta_{ij}\tilde \tau_i$. Let us suppose
that (\ref{eq:induzione_somme_su_j})  be true up to step $R$ and prove it
for the step $R+1$, by summing on $j_{r-R}$.

Since $B^{\tau_R,S_1}_{j_1,\ldots,j_{r-R}}$ has Property A, when
$j_{r-R}$ varies it takes
always the same value (which we will denote by
$B^{\tau_R,S_1}_{\neq}$) unless the index $j_{r-R}$ coincides with at
least one among $j_1,\ldots, j_{r-R-1}$. We thus write
$$
B^{\tau_R,S_1}_{j_1,\ldots,j_{r-R}}=
B^{\tau_R,S_1}_{\neq}+\left(B^{\tau_R,S_1}_{j_1,\ldots,j_{r-R}}
-B^{\tau_R,S_1}_{\neq}\right)\ .  
$$ 
Moreover, since, whenever there exists $l\le r-R-1$ such that $j_l=j_{r-R}$,
$$
1=\frac{\sum_{l=1}^{r-R-1} \delta_{j_l,j_{r-R}}}{\sum_{l=1}^{r-R-1}
  \delta_{j_l,j_{r-R}}} =
\sum_{l=1}^{r-R-1}\frac{\delta_{j_l,j_{r-R}}}{\sum_{l'=1}^{r-R-1}
  \delta_{j_l,j_{l'}}}\ ,
$$
we can also write
$$
B^{\tau_R,S_1}_{j_1,\ldots,j_{r-R}}= B^{
  \tau_R,S_1}_{\neq}+\sum_{l=1}^{r-R-1} \left(B^{
  \tau_R,S_1}_{j_1,\ldots,j_{r-R}} -B^{
  \tau_R,S_1}_{\neq}\right)\frac{\delta_{j_l,j_{r-R}}}{m^{(l)}_{j_1,\ldots,j_{r-R-1}}}\ , 
$$
where, for $l\in\{1,\ldots,r-R-1\}$ the function
$m^{(l)}_{j_1,\ldots,j_{r-R-1}}\equal (\sum_{l'=1}^{r-R-1}\delta_{j_l,j_{l'}})$
counts the number of indices which have the same value as $j_l$.
By summing over $j_{r-R}$ we get
\begin{equation}\label{eq:somma_induzione}
  \begin{split}
\sum_{j_{r-R}}&e^{i2\pi j_1 K\cdot
  \tau^{(1)}/N}\cdots e^{i2\pi j_{r-R} K\cdot
  \tau^{(r-R)}/N} B^{\tau_R,S_1 }_{j_1,\ldots,j_{r-R}} =\\& e^{i2\pi j_1 K\cdot
  \tau^{(1)}/N}\cdots e^{i2\pi j_{r-R-1} K\cdot
  \tau^{(r-R-1)}/N}  N B^{
  \tau_R,S_1 }_{\neq}\sum_{n_{S_1+1}}\delta^{n_{S_1+1}}_{\tau^{(r-R)}\cdot
    K}\\
&\!\!+ \sum_{l=1}^{r-R-1}e^{i2\pi j_1 K\cdot
  \tau_{R+1}^{(1)}(l)/N}\cdots e^{i2\pi j_{r-R-1} K\cdot
  \tau^{(r-R-1)}_{R+1}(l)/N}
 B^{\tau_{R+1}(l),S_1}_{j_1,\ldots,j_{r-R-1}}\\
&\!\!+ \sum_{l=1}^{r-R-1}e^{i2\pi j_1 K\cdot
  \bar\tau_{R+1}^{(1)}(l)/N}\cdots e^{i2\pi j_{r-R-1} K\cdot
\bar  \tau^{(r-R-1)}_{R+1}(l)/N}
\bar B^{\bar \tau_{R+1}(l),S_1}_{j_1,\ldots,j_{r-R-1}}\ ,
  \end{split}
\end{equation}
where
\begin{equation*}
  \begin{split}
B^{\tau_{R+1}(l),S_1}_{j_1,\ldots,j_{r-R-1}}=
\frac{B^{\tau_R,S_1}_{j_1,\ldots,j_{r-R}=j_l}}{m^{(l)}_{j_1,\ldots,j_{r-R-1}}}\ ,\\
\bar B^{\bar \tau_{R+1}(l),S_1}_{j_1,\ldots,j_{r-R-1}}=-
\frac{B^{\tau_R,S_1 }_{\neq}}{m^{(l)}_{j_1,\ldots,j_{r-R-1}}}\ ,
  \end{split}
\end{equation*}
while the collection of vectors $\tau_{R+1}(l)$ and $\bar\tau_{R+1}(l)$
are relied to $\tau_R$ by the following relations:
\begin{equation*}
  \begin{split}
 &   \tau_{R+1}^{(i,j)}(l)=\left\{\begin{array}{cc}
    \tau^{(i,j)}&\mbox{if } 1\le i\le r-R-1,\,i\neq l\\
    \tau^{(i+1,j)}&\mbox{if } r-R\le i\le r-R-1+S_1\\
    \tau^{(l,1)}\cup \tau^{(r-R,1)} &\mbox{if } i=l,\, j=1\\
    \tau^{(l,j')}\vee \tau^{(r-R,j'')} &\mbox{if } i=l,\, 2\le j\le
    S_2(l)+S_2(r-R)-1 \\
    \end{array}\right.\ ,\\
&    \bar\tau_{R+1}^{(i,j)}(l)=\left\{\begin{array}{cc}
    \tau^{(i,j)}&\mbox{if } 1\le i\le r-R-1,\,i\neq l\\
    \tau^{(i+1,j)}&\mbox{if } r-R\le i\le r-R-1+S_1\\
    \tau^{(l,1)}&\mbox{if } i=l,\,j=1\\
    \tau^{(l,\bar j')}\vee \tau^{(r-R,\bar j'')} &\mbox{if } i=l,\,
    2\le j\le S_2(l)+ S_2(r-R)\\
    \end{array}\right.\ ,\\    
\end{split}
\end{equation*}
Here, $2\le j'\le S_2(l)$, $2\le j'' \le S_2(r-R)$ and $2\le \bar j'
\le S_2(l)$, $1\le \bar j''\le S_2(r-R)$, and $\tau^{(l,j)}_{R+1}$,
for $j\ge 2>$ are so chosen that $\min(\supp(\tau_{R+1}^{(l,j)}(l)))\le
\min(\supp(\tau_{R+1}^{(l,k)}(l)))$ if and only if $j<k$, while the
analogous condition holds for $\bar\tau_{R+1}(l)$ (in order that such
collections are admissible).

Notice that $B^{ \tau_R,S_1 }_{\neq}$,
$B^{\tau_{R+1}(l),S_1}_{j_1,\ldots,j_{r-R-1}}$ and $\bar B^{\bar
  \tau_{R+1}(l),S_1}_{j_1,\ldots,j_{r-R-1}}$ are functions of the indices 
$(j_1,\ldots,j_{r-R-1})$ only, since $m^{(l)}_{j_1,\ldots,j_{r-R-1}}$ does.
We observe further that they possess Property A with respect to the set
$(j_1,\ldots,j_{r-R-1})$. In fact, it can be shown directly that
$m^{(l)}_{j_1,\ldots,j_{r-R-1}}$ has such a property, while 
$B^{\tau_R,S_1}_{j_1,\ldots,j_{r-R}=j_l}$ possess it simply because
$B^{\tau_R,S_1}_{j_1,\ldots,j_{r-R}}$ has the corresponding property
with respect to $(j_1,\ldots,j_{r-R})$. As for
$B^{\tau_R,S_1 }_{\neq}$, we recall that, for any
$(j_1,\ldots,j_{r-R-1})$, it is defined as the common value taken by 
$B^{\tau_R,S_1}_{j_1,\ldots,j_{r-R}}$ for all $j_{r-R}\neq j_i$, for
$i<r-R$. So, by fixing arbitrarily $n<r-R-1$ indices among
$(j_1,\ldots,j_{r-R-1})$  and taking for the remaining indices
$j_{i_{n+1}}=\ldots=j_{i_{r-R-1}}$ the common value $\bar \jmath$, one has
$$
B^{\tau_R,S_1 }_{\neq}= B^{\tau_R,S_1
}_{j_1,\dots,j_{r-R-1},j_{r-R}}\ , \quad\mbox{ for any } j_{r-R}\neq
  j_{i_1},\ldots, j_{i_n},\bar \jmath\ .
  $$
  We fix $j_{r-R}$ and let $\bar \jmath$ vary: by Property A for
  $B^{\tau_R,S_1}_{j_1,\ldots,j_{r-R}}$,  $B^{\tau_R,S_1 }_{\neq}$
  takes the same value for all $\bar \jmath \neq j_{i_1},\ldots,
  j_{i_n}, j_{r-R}$. Then we change $j_{r-R}$, thus showing that the
  same holds true for all $\bar \jmath \neq j_{i_1},\ldots,
  j_{i_n}$, i.e., exactly Property A with respect to
  $(j_1,\ldots,j_{r-R-1})$.

This way we have completed the proof of the inductive
hypothesis (\ref{eq:somma_induzione}) at step $R+1$, and so
equation (\ref{eq:btau}). 

In order to prove estimate (\ref{eq:btau_stima}), we use a graphical
tool to keep track of the number of addenda which contribute to any
collection $\tau$.

\begin{figure}
  \begin{center}
  \includegraphics[width=0.8\textwidth]{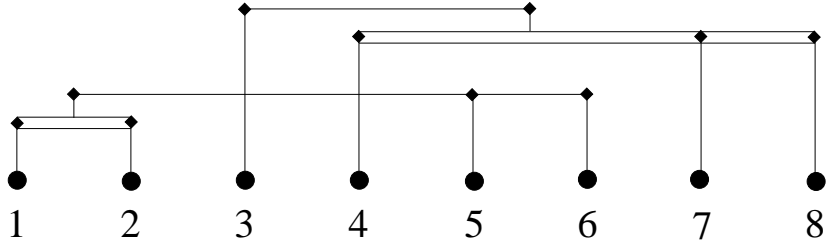}
  \end{center}
  \caption{\label{fig:grafico1} Example of graph of the first type.}
\end{figure}

We associate to any $\tau$ a graph in the following way: we draw $r$
points, corresponding to the indices $\{1,\ldots,r\}$ in this order,
and connect them in such a way that two sites belong to the support of the same
$\tau^{(i)}$ if and only if there exists at least a line joining them
and that they belong to the support of the same $\tau^{(i,j)}$
if and only if they are joined by a double line. The example of
Figure~\ref{fig:grafico1} displays a case in which $r=8$ and
$\supp(\tau^{(1,1)})=\{1,2\}$, 
$\supp(\tau^{(1,2)})=\{5\}$, $\supp(\tau^{(1,3)})=\{6\}$,
$\supp(\tau^{(2,1)})=\{3\}$, $\supp(\tau^{(2,2)})=\{4,7,8\}$. We point
out that the correspondence between the collections $\tau$ and the
graph is biunivocal, as the order of the 
$\tau^{(i,j)}$ is univoquely assigned.

\begin{figure}
  \begin{center}
    \subfigure{\includegraphics[width=0.45\textwidth]{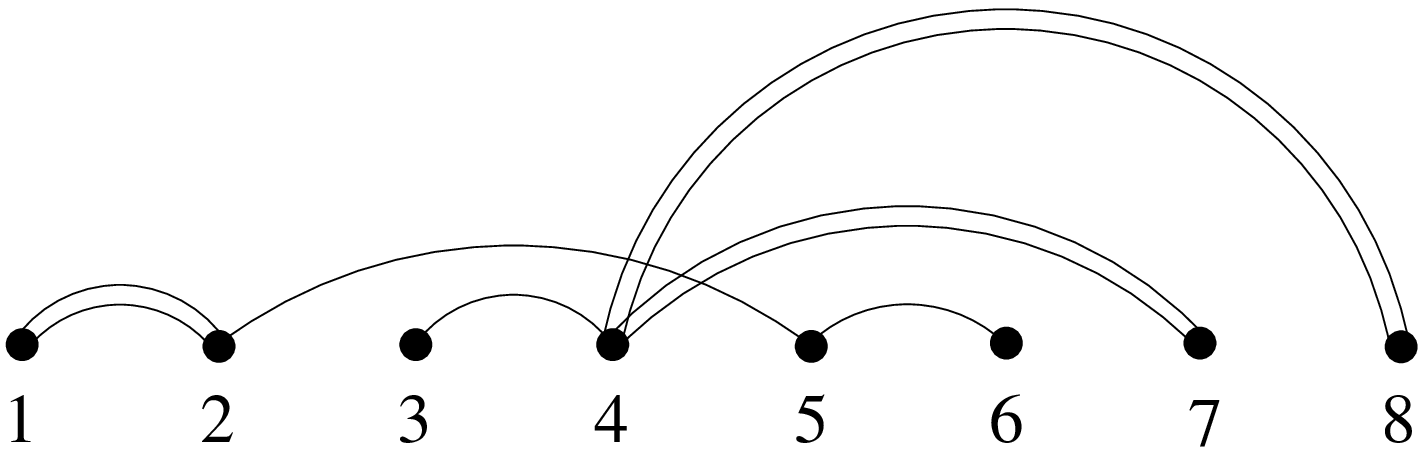}
      \quad \includegraphics[width=0.45\textwidth]{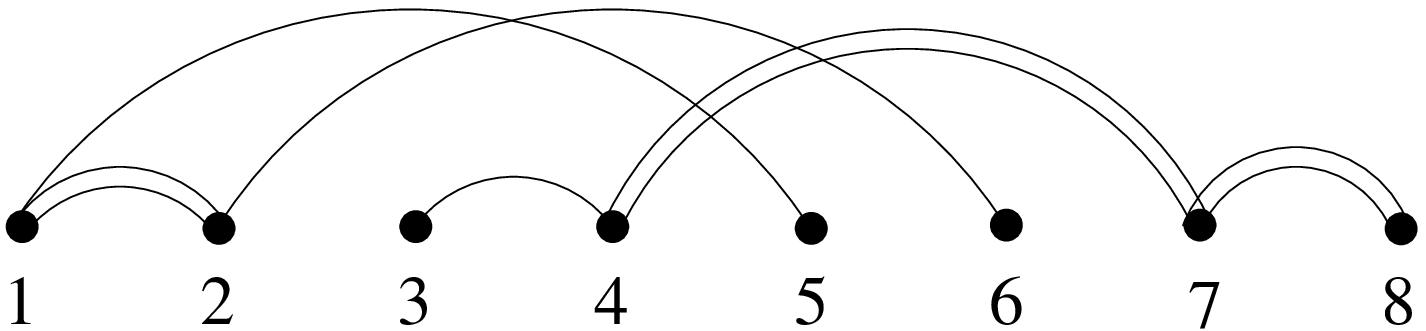}}
  \end{center}
  \caption{\label{fig:grafico2} Two different example of graphs of the
    second type corresponding to a single graph of the first type,
    namely, the graph in Figure~\ref{fig:grafico1}.}
\end{figure}

The term corresponding to a given  $\tau$ can come from different
terms in the sum over $(j_r,\ldots,j_1)$, as we illustrate now, by
constructing another type of graph, where the only difference with
respect to the previous one lies in the
form of the lines. For any site $r-R$, for $0\le R\le r$, three
alternatives are possible: 
\begin{itemize}
\item no line  pointing left comes out of the site;
    \item one double line pointing left comes out of the point
      and joins it with one site 
      on its left;
\item one simple line pointing left starts from the point and ends on one point on its left.
\end{itemize}
Such options correspond to the first, to the second
and to the third term at the r.h.s. of (\ref{eq:somma_induzione}),
respectively. This 
way we associate with a bijection to any graph one single term coming
from the sums on $(j_r,\ldots,j_1)$. Moreover, every graph of this
kind is associated to one and only one graph of the previous type via
the prescription that in the latter two points are joined by a line (be
it a simple or a double line) if and only if there exists a line, or a set
of lines, of the same kind which connects them continuously in the
former. We point out that the relation between the two types of graph
is \emph{not} biunivocal. In Figure~\ref{fig:grafico2} we show two
examples of terms with different graphs of the second type,
corresponding to the same $\tau$ as in Figure~\ref{fig:grafico1}. 

We have reduced the estimate (\ref{eq:btau_stima}) basically to the
count of the number of graphs of the second type giving the same graph
of the first kind. Indeed, let us remark the following facts, which
can be checked from (\ref{eq:somma_induzione}):
\begin{enumerate}
\item all terms corresponding to a given  $\tau$ depend on 
  $B_J$ only for $J\in \mathcal J(\tau)$;
  \item in the graphs of the second type, if two or more sites are
    connected by a chain of double lines (in such a case we will call
    the maximal set of sites forming one of such groups a \emph{double
    chain}), a possible 
    simple line joining a point 
    of the double chain to a point on the left of the group can start only
    from the leftmost point in the double chain;
    \item any line starting from a point  $m$ in a graph of the second
      type and ending on the point $l<m$ implies that the
      corresponding term is divided by
      $$
\sum_{j=1}^{m-1} \delta^=_{j,l}\ ,
$$
with $\delta^=_{j,l}=1$ if $j$ and $l$ belong to the same double chain (or coincide), 0 otherwise.
\end{enumerate}

Due to item 3 all terms linking a point, through a simple
line,  to a double chain on the left of the point count
algebraically as one single term, whereas by item 2 the same happens
for the connection between two double chains through a simple
line. Still by item 3, all terms forming a double chain count
algebraically as one. In order to estimate
$c^{\tau,S_1}_{\tilde \tau}$ it is then sufficient to count the number
of possible connections between the different double chains, in such a way
that each double chain is connected with another one on its left. For any
$\tau^{(i)}$, we have $S_2(i)$ double chains, which can be connected
in $(S_2(i)-1)!$ ways. This completes the estimate 
(\ref{eq:btau_stima})
%
and concludes the proof of Lemma~\ref{lemma:somme_su_j}.

\vskip 2. ex
\noindent
From this lemma, and in particular from equation (\ref{eq:btau}),
Lemma~\ref{lemma:somma_scambiabili} immediately follows. In order to
prove Lemma~\ref{lemma:indipendenti}, instead, we use
(\ref{eq:btau_stima}) together with Lemma~\ref{lemma:misura}. In our
case this gives, for
$\tau\in \bar \Tc$, 
$$
c^{\tau,S_1}_{\tilde \tau}\le\frac{K^{2s}}{N\beta^s} \sqrt{2s!}
\prod_{i=1}^{S_1} (S_2(i)-1)!\ .
$$
For fixed $\tau^{(i)}$ and $S_2(i)$, the number $\mathfrak
m_{i,S_s(i)}$ of collections
$(\tau^{(i,1)},\ldots, \tau^{(i,S_2(i))})$ giving the same
$\tau^{(i)}$ is
\begin{equation*}
  \begin{split}
\mathfrak m_{i,S_2(i)}=&\sum_{\mathfrak n_{i,1}=1}^{\mathfrak n_i-S_2(i)+1} \frac{(\mathfrak
  n_i-1)!}{(\mathfrak n_i-\mathfrak n_{i,1})!(\mathfrak n_{i,1}-1)!}
\cdots \\
&\sum_{\mathfrak n_{i,j}=1}^{\mathfrak n_i-\mathfrak n_{i,1}
  -\cdots - \mathfrak n_{i,j-1}- S_2(i)+j} \frac{(\mathfrak
  n_i-\mathfrak n_{i,1}-\cdots - \mathfrak n_{i,j-1}- 1)!}{(\mathfrak
  n_i-\mathfrak n_{i,1}-\cdots-\mathfrak n_{i,j})!(\mathfrak
  n_{i,j}-1)!} \cdots\\
&\sum_{\mathfrak n_{i,S_2(i)-1}=1}^{\mathfrak n_i-\mathfrak n_{i,1}
  -\cdots - \mathfrak n_{i,S_2(i)-2}-1} \frac{(\mathfrak
  n_i-\mathfrak n_{i,1}-\cdots - \mathfrak n_{i,S_2(i)-2}- 1)!}{(\mathfrak
  n_i-\mathfrak n_{i,1}-\cdots-\mathfrak n_{i,S_2(i)-1})!(\mathfrak
  n_{i,S_2(i)-1}-1)!}
  \end{split}
\end{equation*}
Due to the binomial expansion, we have
$$
\sum_{j=1}^{m-1}\frac{(m-1)! l^{m-1-j}}{(m-j)!(j-1)!} \le
\frac{(l+1)^{m-1}}l\ .
$$
By applying repeatedly this formula, with $m=\mathfrak n_i-\mathfrak n_{i,1}
  -\cdots - \mathfrak n_{i,j-1}$ e $j=\mathfrak n_{i,j}$, we get
  $$
\mathfrak m_{i,S_2(i)}\le \frac{S_2(i)^{\mathfrak
    n_i-1}}{(S_2(i)-1)!}\ ,
$$
so that, for some suitable $C\ge 0$,
$$
\mathfrak m_i\equal\sum_{S_2(i)=1}^{\mathfrak n_i}\mathfrak m_{i,S_2(i)} (S_2(i)-1)!\le
C^{\mathfrak n_i} \frac{(\mathfrak n_i-1)!}{\mathfrak n_i}\ .
$$

We then sum over all possible
$(\tau^{(1)},\ldots,\tau^{(S_1)})$, obtaining
\begin{equation*}
  \begin{split}
&\sum_{(\tau^{(1)},\ldots,\tau^{(S_1)})\in \mathcal T^{2s}}
\prod_{i=1}^{S_1}\mathfrak m_i \mathfrak n_i= 2^{2s} \sum_{\mathfrak
  n_1=1}^{2s-S_1+1} \frac{(2s-1)!\mathfrak m_1\mathfrak n_1}{(2s-\mathfrak
  n_1)!(\mathfrak n_1-1)!} 
\cdots \\
&\qquad\qquad\sum_{\mathfrak n_1=1}^{2s-\mathfrak n_1
  -\cdots - \mathfrak n_{i-1}- S_1+i} \frac{(2s-\mathfrak n_1-\cdots -
  \mathfrak n_{i-1}- 1)!\mathfrak m_i\mathfrak n_i}{(2s-\mathfrak
  n_1-\cdots-\mathfrak n_i)!(\mathfrak  n_i-1)!} \cdots\\
&\qquad\qquad\sum_{\mathfrak n_{S_1-1}=1}^{2s-\mathfrak n_1
  -\cdots - \mathfrak n_{S_1-2}-1} \frac{(2s-\mathfrak n_1-\cdots -
  \mathfrak n_{S_1-2}- 1)!\mathfrak m_{S_1-1} \mathfrak m_{S_1}
  \mathfrak n_{S_1-1} \mathfrak n_{S_1}}{(2s-\mathfrak n_1-\cdots-\mathfrak n_{S_1-1})!(\mathfrak
  n_{S_1}-1)!}\ .
  \end{split}
\end{equation*}
Since $\mathfrak m_i\mathfrak n_i/(\mathfrak n_i-1)!\le C^{\mathfrak
  n_i}$ and $ \sum_i\mathfrak n_i=2s$, we get
$$
\sum_{(\tau^{(1)},\ldots,\tau^{(S_1)})\in \mathcal T^{2s}}
\prod_{i=1}^{S_1}\mathfrak m_i \mathfrak n_i \le C_1^{2s} (2s)!\ .
$$
A last sum over $S_1$ and a suitable choice of constants bring us to
the proof of the statement of Lemma~\ref{lemma:indipendenti}
for $(\tau^{(1)},\ldots,\tau^{(S_1)})\in \bar\Tc$. The case of
$(\tau^{(1)},\ldots,\tau^{(S_1)})\in \bar\Tc^c$ is dealt with in a
completely analogous way.

\end{document}